\documentclass[twocolumn,showpacs,superscriptaddress,nofootinbib]{revtex4-1}
\usepackage{graphicx}
\usepackage{subfigure}
\usepackage{epsfig}
\usepackage{epstopdf}
\usepackage{color}
\usepackage{amsmath}

\newcommand{\tcb}{\textcolor{blue}}
\newcommand{\tcr}{\textcolor{red}}

\newcommand{\rbm}[1]{{\color{red}\bf [Robb: #1]}}

\newcommand{\be}{\begin{equation}}
\newcommand{\ee}{\end{equation}}
\newcommand{\ba}{\begin{eqnarray}}
\newcommand{\ea}{\end{eqnarray}}
\newcommand{\beq}{\begin{equation}}
\newcommand{\eeq}{\end{equation}}
\newcommand{\beqa}{\begin{eqnarray}}
\newcommand{\eeqa}{\end{eqnarray}}

\usepackage[normalem]{ulem}

\begin{document}

\title{Thermodynamics of Dyonic NUT Charged Black Holes with Entropy as Noether Charge}

\author{Niloofar Abbasvandi}
\email{niloofar.abbasvandi@uwaterloo.ca}
\affiliation{Boxbrite Technologies Inc., 2 Florapine Rd. Floradale, Ontario, Canada,  N0B 1V0}
\affiliation{Department of Physics and Astronomy, University of Waterloo, Waterloo,
Ontario, Canada, N2L 3G1}
\author{Masoumeh Tavakoli}
\email{tavakoli.phy@gmail.com}
\affiliation{Department of Physics and Astronomy, University of Waterloo, Waterloo,
Ontario, Canada, N2L 3G1}
\affiliation{Department of Physics, Isfahan University of Technology, Isfahan, Iran}
\author{Robert B. Mann}
\email{rbmann@uwaterloo.ca}
\affiliation{Department of Physics and Astronomy, University of Waterloo, Waterloo,
Ontario, Canada, N2L 3G1}

\date{\today}

\begin{abstract}
We investigate the  thermodynamic behaviour of Lorentzian Dyonic Taub-NUT Black Hole spacetimes.  
We consider two possibilities in their description: one in which their entropy is interpreted to be one quarter of the horizon area (the horizon entropy), and another in which the Misner string also contributes to the entropy (the Noether charge entropy).  
We find that there can be as many as three extremal black holes (or as few as zero) depending on the choice of parameters,
and that the dependence of the free energy on temperature -- and the
 resultant phase behaviour -- depends very much on which of these situations holds.     
Some of the phase behaviour we observe holds regardless of which interpretation of the entropy holds.  However 
another class of phase transition structures occurs only if the Noether charge interpretation of the entropy is adopted.
\end{abstract}

\maketitle

\section{introduction}

Once referred to as a ``counterexample to almost anything" \cite{1967rta1.book..160M},  Taub-NUT spacetime was generally regarded as an unphysical solution to Einstein gravity, since it had  rotating string like singularities (Misner strings) and  regions of closed timelike curves (CTCs) in their vicinity.  Its Euclidean version became the preferable form for the metric, the solution being interpreted as a gravitational instanton \cite{EGH}.  Its thermodynamic properties were later interpreted in this context
\cite{Hawking:1998ct,Chamblin:1998pz,Noether,Mann:1999pc,Emparan:1999pm,Mann:1999bt,Johnson:2014xza,Johnson:2014pwa}, the key point being that a periodic identification of the (Euclidean) time coordinate is made so that the Misner string singularity is removed \cite{Misner}. Apart from the consequence that the Lorentzian version of the spacetime has  CTCs everywhere, the maximal extension of the spacetime becomes problematic  \cite{Misner, hawking1973large,hajicek1971causality}.

Recently there has been a revival of interest in the Lorentzian Taub-NUT (LTN) spacetime 
\cite{Clement:2015cxa,Clement:2015aka,Clement:2016mll,Clement:2017kas,Kubiznak:2019yiu,Bordo:2019tyh,Durka:2019ajz,Ballon:2019uha,Bordo:2019rhu}.  The primary reason for this was the recent demonstration that LTN spacetime is geodesically complete and that freely falling observers do not experience causal pathologies \cite{Clement:2015cxa,Clement:2015aka}.  Although the latter situation does not hold for other (non-geodesic) observers, it has been argued that spacetime geometry would be deformed by
the  back-reaction of these  accelerations so that chronology is ultimately be preserved \cite{Clement:2015cxa}.  Provided this is the case, there is no apparent obstruction toward considering LTN as a physically admissible solution to Einstein gravity.   Indeed, the shadows of rotating LTN black holes have been constructed \cite{Abdujabbarov:2012bn,Grenzebach:2014fha}, anticipating possible observational constraints on these objects.

Our interest in this paper is in the thermodynamic behaviour of LTN black holes.  Work on this originated with the de Sitter LTN  \cite{Mann:2004mi} and on the tunnelling method for computing black hole temperature \cite{Kerner:2006vu}.   More recently there have been more complete studies in which the free energy has been calculated \cite{Kubiznak:2019yiu,Durka:2019ajz} and a formulation of the laws of NUT-charged black hole mechanics \cite{Bordo:2019tyh,Durka:2019ajz,Ballon:2019uha,Bordo:2019rhu} were derived\footnote{These laws have since been used to investigate weak cosmic censorship for NUT-charged black holes \cite{Yang:2020iat,Feng:2020tyc}.}, generalizing the original approach \cite{Bardeen}. Unlike the Euclidean case, this approach yields a first law of full cohomogeneity.  The Misner strings can be asymmetrically distributed along the north-south polar axes, and the  gravitational Misner charges encode their strengths.
 
Despite these successes, there remains an ambiguity in this approach, namely an identification of the entropy \cite{Kubiznak:2019yiu}.  The Noether charge method applied to the Euclidean solution \cite{Noether} yields an entropy $S_{{\cal N}}$ that is a combination of contributions from the horizon area and the Misner strings.  The temperature $T$ is given via either the tunnelling method \cite{Kerner:2006vu} or by standard Wick-rotation arguments and is the surface gravity of the black hole.  A new pair of conjugate variables $(\psi,N)$ appear that ensure full cohomogeneity of the first law.  All thermodynamic quantities are finite for all finite values of the NUT charge $n$, and  have a smooth limit as the $n\to 0$. 

However it was subsequently argued  \cite{Bordo:2019tyh} that the surface gravity of the black hole and its conjugate areal quantity should respectively correspond to the temperature and entropy of the LTN black hole, with an additional conjugate pair of variables $(\psi^\prime_{\textsf{N/S}},N^\prime_{\textsf{N/S}})$ corresponding to  the surface gravity of the Misner strings  and a conjugate Misner charges, the ${\textsf{N/S}}$  corresponding to the north/south polar axes.    This approach is more geometrically intuitive, but has the feature that one of $\psi^\prime_{\textsf{N/S}}$ diverges at some finite value of $n$, and has no smooth $n\to 0$ limit if the Misner strings are symmetrically distributed.  It is likewise unclear if the $\psi^\prime_{\textsf{N/S}}$ should be interpreted as temperatures associated with the strings (in which case 
$N^\prime_{\textsf{N/S}}$ are the corresponding string entropies) or not \cite{Bordo:2019tyh}.

Furthermore, the choice of thermodynamic potentials for the NUT charged black hole was recently shown to be ambiguous when both electric and magnetic charge are present \cite{Ballon:2019uha}.  Two possible versions of the thermodynamic first law can be formulated depending on how these charges are defined. Both the magnetic and electric charges depend on the radius of the sphere over which the field strength and its dual are integrated via Gauss' law.  One can either take the magnetic charge to be the value at infinity and the electric charge to be that at the horizon, or vice-versa.  In both cases a first law of full cohomogeneity in all variables is obtained, but the thermodynamic NUT charges differ, related to each other by
electromagnetic duality \cite{Ballon:2019uha}.  

A somewhat analogous situation holds for the choice of entropy.  The two approaches are connected by the relation
\be\label{S-rel}
 S_{{\cal N}} =  \frac{A}{4}  + \frac{\psi^\prime_{\textsf{N}} N^\prime_{\textsf{N}}  +  \psi^\prime_{\textsf{S}} N^\prime_{\textsf{S}}}{T}
\ee
where $A$ is the area of the black hole.  If  $\psi^\prime_{\textsf{N}} = \psi^\prime_{\textsf{S}} \propto T$, a situation that would arise under analytic continuation of periodic identification of the temperature \cite{Misner,Mann:2004mi} then the Noether charge entropy would equal the entropy from the horizon plus the total entropy
$ N^\prime_{\textsf{N}}  +  N^\prime_{\textsf{S}}$ of the strings \cite{Bordo:2019tyh}, if the latter indeed can be regarded as string entropies.

It is the purpose of this paper to study the thermodynamics of charged LTN black holes under these two interpretations -- one in which the entropy is taken to be $S_{{\cal N}}$ and the other in which the entropy 
is taken to be $S_+ = \frac{A}{4}$.   Previous work on this subject \cite{Ballon:2019uha}  considered only the latter interpretation.  It is our purpose here to understand what the thermodynamic implications are of considering the entropy to the the Noether charge entropy $S_{{\cal N}}$, in accord with all other approaches to black hole thermodynamics.

We shall work in the context of Black Hole Chemistry \cite{Kubiznak:2016qmn}. This approach, in which
the cosmological constant $\Lambda$ is regarded as a thermodynamic variable
\cite{Creighton:1995au,Caldarelli:1999xj,Padmanabhan:2002sha,Kastor:2009wy} 
corresponding to a pressure $P = -\frac{\Lambda}{8 \pi} = \frac{3}{8 \pi l^2}$, has proven to be very fruitful.   The Hawking-Page transition for AdS black holes
\cite{Hawking and Page}  can be reinterpreted in terms of a first-order liquid/solid phase transition
\cite{Kubiznak:2014zwa}.  Many new phenomena appear, including van der Waals phenomena \cite{Kubiznak:2012wp},  re-entrant phase transitions \cite{Altamirano:2013ane,Frassino:2014pha}, black hole triple points \cite{Altamirano:2013uqa}, polymer-like transitions \cite{Dolan:2014vba}, superfluid phase transitions  \cite{Hennigar:2016xwd}, repulsive black hole microstructure \cite{Wei:2019uqg}, and more  \cite{Kubiznak:2016qmn}.

This task is somewhat complicated since the conserved electric and magnetic charges depend on the 
radius of the 2-sphere that encloses the black hole.  If one requires the electromagnetic vector potential to
vanish at the horizon \cite{Johnson:2014pwa, Awad:2005ff,Dehghani:2006dk}, then the electric and magnetic charges are no longer independent, and one can
take the conserved electric charge to that given as $r\to\infty$.  As a consequence of this, the first law
of thermodynamics no longer has full cohomogeneity.  Only one of the electric/magnetic charges appears in the first law, even though the  LTN black hole has both types of charges.   Furthermore, it has been shown that this 
constraint is not necessary:   all the parameters of the solution can be varied independently varied
provided that one charge is an asymptotic charge ($r\to\infty$) and the other is evaluated on the horizon
($r\to r_+$) \cite{Ballon:2019uha}.  We shall consider all three scenarios: asymptotic electric charge, asymptotic magnetic charge, and the standard `constrained' thermodynamics.

Within each scenario we shall consider a variety of ensembles to see what kinds of phase behaviour possible in each. Along with the Noether-charge entropy $S_{{\cal N}}$, the LTN solution has two 
conjugate thermodynamic NUT charge/potential pairs, $(N_{{\cal N}{\textsf{N}}} ,\psi_{{\cal N}{\textsf{N}}})$ and $(N_{{\cal N}{\textsf{S}}} ,\psi_{{\cal N}{\textsf{S}}})$ associated with each polar axis.  For simplicity we shall choose these to be equal, referring to them as  
$(N_{{\cal N}} ,\psi_{{\cal N}})$.  We therefore consider the following
ensembles
\begin{enumerate}
\item Fixed electric and magnetic charges, fixed $N_{{\cal N}}$
\item Fixed electric and magnetic charges, fixed $\psi_{{\cal N}}$
\item Fixed electrostatic potential, fixed magnetic charge, fixed $N_{{\cal N}}$
\item Fixed electrostatic potential, fixed magnetic charge, fixed $\psi_{{\cal N}}$
\item Fixed magnetostatic potential, fixed electric charge, fixed $N_{{\cal N}}$
\item Fixed magnetostatic potential, fixed electric charge, fixed $\psi_{{\cal N}}$
\end{enumerate}
for each of the three scenarios for defining charge.   Furthermore, the relation \eqref{S-rel} implies that
each of these ensembles has a counterpart in a setting where $S_+$ is regarded as the entropy
with corresponding charge/potential pair, $(N_+,\psi_+)$,  with  fixed $N_{{\cal N}}$ corresponding to
fixed $\psi_+$ and vice-versa.  This latter situation is equivalent to fixing the NUT charge $n$, and its thermodynamics
has been given some study previously \cite{Ballon:2019uha}.  

We shall therefore concentrate on the ensembles with
fixed $\psi_{{\cal N}}$. We find  several new phase phenomena in this case that we refer to as the fractured cusp, snapping cusp, zig-zag, and
double swallowtail structures in plots of the free energy as a function of temperature.




\section{The Charged Lorentzian Taub NUT Solution}

The charged LTN metric is \cite{Johnson:2014pwa,AlonsoAlberca:2000cs}

\begin{equation}\label{metric}
    ds^2=-f[dt+2n \cos\theta d\phi]^2+\frac{dr^2}{f}+
    (r^2+n^2)(d\theta ^2+ \sin^2\theta d\phi^2)
\end{equation}

where 
\be\label{Apot}
A=-h(dt + 2 n \cos\theta d\phi)
\ee
is the electromagnetic vector potential.  The functions $f$ and $h$ are
\begin{align}\label{met-f}
f &=\frac{r^2-2 m r -n^2+4 n^2 g^2 +e^2}{r^2+n^2}-\frac{3n^4-6n^2r^2-r^4}{l^2(r^2+n^2)} \\
h &=\frac{er}{r^2+n^2}+\frac{g(r^2-n^2)}{r^2+n^2}
\label{met-h}
\end{align}
with $n$  the NUT parameter, $m$ the mass parameter, and $e$ and $g$ the respective electric and magnetic charge parameters.  The thermodynamic pressure is
\be
P = -\frac{\Lambda}{8\pi}  =  \frac{3}{8\pi l^2} 
\ee
and 
\be
V = \frac{4\pi r_+}{3}\left(r_+^2 + 3n^2\right)
\ee
is its conjugate thermodynamic volume \cite{Kubiznak:2019yiu}.

The  mass and angular momentum can be computed via conformal completion methods 
\cite{Ashtekar:1999jx,Das:2000cu}, yielding for the mass
\begin{align}\label{Mass}
M &= m \\
&= \frac{e^2l^2-3n^4+6n^2r_+^2+r_+^4+l^2((4g^2-1)n^2+r_+^2)}{2l^2r_+} \nonumber
\end{align}
where $f(r_+)=0$; the angular momentum vanishes \cite{Ballon:2019uha}. The conserved electric and magnetic
charges are respectively given by
\begin{align}\label{charge-e}
&q_e=\frac{1}{4\pi}\int_{S^2} *F = \frac{e(r^2-n^2)- 4 g r n^2}{r^2+n^2} \\
&q_m=\frac{1}{4\pi}\int_{S^2} F = -2n \frac{g(r^2-n^2) + er }{r^2+n^2}  
\label{charge-m}
\end{align}
and depend on the radius of the sphere over which the integration is performed \cite{Ballon:2019uha}. Their
asymptotic values are
\begin{equation}\label{charges2}
  Q =  \lim_{r\to\infty} q_ e = e  \qquad
  Q_m=  \lim_{r\to\infty} q_ m  = -2gn 
  \end{equation}

The electrostatic potential $\phi$ is obtained from calculating the difference between the values of $-\xi.A$ on the horizon and infinity \cite{Ballon:2019uha} 
 \begin{equation}\label{phidiff}
 \phi= -(\xi.A|_{r= r_+} -\xi.A|_{r=\infty}) = \frac{e r_+- 2g n^2}{r_+^2+n^2}
\end{equation}
and 
is the conserved electric charge.

The temperature associated with the surface gravity at the horizon is
\begin{equation}\label{Temp}
 T=\frac{f'_+}{4\pi}=\frac{1}{4\pi r_+}\left(1+\frac{3(r_+^2+n^2)}{l^2}-\frac{e^2+4 n^2 g^2}{r_+^2+n^2}\right)
\end{equation}
and 
\be\label{SBH}    
S_+ = \pi (r_+^2+n^2)   
\ee
is the contribution to the entropy from the horizon of the black hole.

To compare the 2 choices of entropy and thermodynamic NUT charge we
must have
\begin{equation}\label{2flaw}
TdS_+  +  \psi_+ dN_+ = TdS_{{\cal N}} + \psi_{{\cal N}} dN_{{\cal N}}
\end{equation}
so that the first law holds for both choices of the entropy, where 
\be \label{psi+}
\psi_+ = \frac{1}{8\pi n}
\ee
is the thermodynamic potential
corresponding to the thermodynamic NUT charge $N_+$.  As noted above, this latter quantity is contingent upon the choice of electric and magnetic charge. For the choice of metric 
\eqref{metric} the potential $\psi_{-} = 0$  \cite{Bordo:2019tyh}.  We note also that $\psi_+$ diverges as $n\to 0$, making the thermodynamic interpretation of this potential less than clear. 

The Noether charge entropy $S_{{\cal N}}$   contains contributions from both the horizon and the Misner string, with $N_{{\cal N}}$  and $\psi_{{\cal N}}$  the corresponding thermodynamic NUT charge and conjugate potential.  To relate these two we can write
\be
N_+ = N_{{\cal N}} X   \qquad   X \psi_+ = \psi_{{\cal N}}
\ee
where $X$ is a function of the  parameters $(r_+,n,l,e,g)$. Upon insertion into \eqref{2flaw} we obtain
\begin{align}
TdS_+ + \psi_+ dN_+ &= TdS_+ + \frac{\psi_{{\cal N}}}{X} d(X N_{{\cal N}}) \nonumber\\
&=  TdS_+ + \frac{N_{{\cal N}} \psi_{{\cal N}}}{X} d(X) + 
 {\psi_{{\cal N}}} d N_{{\cal N}} \nonumber \\
& = TdS_{{\cal N}} + \psi_{{\cal N}} d N_{{\cal N}}
\end{align}
yielding 
\be\label{XNpsi}
X = \frac{N_{{\cal N}} \psi_{{\cal N}}}{T}    \Rightarrow  N_+ = k \frac{N^2_{{\cal N}} \psi_{{\cal N}}}{T}   \qquad  \psi_+ =  \frac{T}{k N_{{\cal N}}}   
\ee
and
\be\label{SNS+}
S_{{\cal N}}  = S_+ + \frac{N_+ \psi_+}{T}   \leftrightarrow   S_+  =  {S_{{\cal N}}  - \frac{N_{{\cal N}} \psi_{{\cal N}}}{T}  }
\ee
where $k$ is a non vanishing numerical constant whose choice is arbitrary.  We shall choose
$k=2$ henceforth to agree with previous conventions  \cite{Kubiznak:2019yiu,Durka:2019ajz}.

The Noether charge approach ascribes the total thermodynamic entropy as having
contributions from both the horizon and the Misner string \cite{Noether}.  This latter quantity
depends on $N_+$, which itself depends on the definition of electric and magnetic charge.
From \eqref{XNpsi} it is easy to see that $N_{{\cal N}}$ does not depend on these definitions,
but that $\psi_{{\cal N}}$ does.  Consequently the Noether charge entropy likewise depends on this choice.
 
We note also that   $S_{{\cal N}}$ diverges as the black hole approaches extremality.  This is not possible if $e=g=0$.  But for any nonzero $\{e,g\}$ the temperature of the black hole vanishes for
\be
 r^{\textrm{ext}}_+ = \frac{l}{\sqrt{6}}\sqrt{\sqrt{1+ 12\frac{e_+^2}{l^2} + 48 g_+^2 n^2}- \left(1+6\frac{n^2}{l^2}\right) }
\ee
 where the notation $e_+$ and $g_+$ indicate that $e$ or $g$ could depend on $r_+$ given 
the definitions \eqref{charge-e} and \eqref{charge-m}.  We see that
sufficiently small values of $n$ exist for which $ r^{\textrm{ext}}_+$ is real and positive and hence
for which $S_{{\cal N}}$ diverges.  For fixed values of $e$ and $g$,  this implies there is a threshold value of the pressure
\be
P_t = \frac{4 g^2 n^2 +e^2 - n^2}{8\pi n^4}
\ee
that determines whether or not the temperature can vanish.  For $P < P_t$, $T > 0$ for all values of $r_+$.

The diverging behaviour of  $S_{{\cal N}}$ is  an obvious consequence of \eqref{SNS+}, and is perhaps the best reason for regarding the horizon
entropy as the entropy of the LTN black hole.   However it is the purpose of this paper to explore the physical implications of interpreting the entropy of this object to come from both the horizon and the Misner string.  If indeed the Misner string has gravitational degrees of freedom, these likewise could contribute to the entropy of the black hole.  This is the approach taken in the Euclidean case, and it is our purpose to understand the implications of this in the Lorentzian sector.

We also note that for any given value of $n \neq 0$ there exist values of $\{e,g\}$ so that the LTN black hole is sub-extremal.

In subsequent sections we shall consider the follow definitions of electric and magnetic 
charge  \cite{Ballon:2019uha}:
\begin{enumerate}
\item Horizon magnetic charge, given by \eqref{charge-m} with $r=r_+$ and 
electric charge   at infinity, given by \eqref{charge-e} with $r\to \infty$.
\item Horizon electric charge, given by \eqref{charge-e} with $r=r_+$ and 
magnetic charge   at infinity, given by \eqref{charge-m} with $r\to \infty$.
\item A constrained thermodynamics  where the electric and magnetic charges
are related by imposing the constraint that the electromagnetic potential 
$A$ in \eqref{Apot} vanishes at the horizon \cite{Johnson:2014pwa, Awad:2005ff,Dehghani:2006dk}.
 \end{enumerate} 
 
Before proceeding, in considering the phase  transition behaviour for various ensembles,
the relation \eqref{SNS+} implies that 
fixed $N_{\cal N}$ in scenarios where $S_{\cal N}$ is regarded as the entropy corresponds to fixed $\psi_+$ in scenarios where $S_{+}$ is regarded as the entropy.   Likewise,
fixed $\psi_{\cal N}$ in scenarios where $S_{\cal N}$ is regarded as the entropy corresponds to fixed $N_+$ in scenarios where $S_{+}$ is regarded as the entropy.    
In what follows we shall adopt the perspective that $S_{\cal N}$ is regarded as the entropy,
and will comment where relevant as to what distinctions arise if $S_{+}$ is regarded as the entropy.

\section{Case 1: Horizon Magnetic Charge}

For this first case we consider horizon magnetic charge $Q_m^{(+)}\equiv q_m(r=r+)$, yielding
from \eqref{charge-m}  
\begin{equation}\label{case 1}
    Q_m^{(+)} = \frac{-2n(er_+ + g[r_+^2-n^2]) }{r_+^2+n^2} 
    \qquad Q_e=e
\end{equation}
where the electric charge is   given by \eqref{charges2}. 
The corresponding potentials are given by :
\begin{align}\label{potentials}
&\phi^{(1)}_m= \frac{-n(2 g r_+ + e) }{r_+^2+n^2} \\
&\phi^{(1)}_e=\frac{-2gn^2 + e r_+ }{r_+^2+n^2}
\nonumber 
\end{align}

\begin{widetext}
In this case the first law incorporating horizon entropy $S_+$ is
\be\label{flaw+}
dM = TdS_+ + \psi_+ dN^{(1)}_+ + VdP +  \phi^{(1)}_e dQ_e  +  \phi^{(1)}_m dQ^{(+)}_m
\ee
provided  \cite{Ballon:2019uha}
 \begin{align}
&N^{(1)}_+=-\frac{4\pi n^3}{r_+}\left(1+\frac{3(n^2-r_+^2)}{l^2} 
  +\frac{(r_+^2-n^2)(e^2+4ger_+)}{(r_+^2+n^2)^2}-\frac{4n^2g^2(3r_+^2+n^2)}{(r_+^2+n^2)^2}\right) \quad .
\label{N1+}  
 \end{align}

Using \eqref{XNpsi} and \eqref{SNS+}, we obtain
\begin{align}
S^{(1)}_{{\cal N}} & =\frac{
(l^2(r_+^4+4n^2r_+^2 -n^4 )e^2+8l^2n^2r_+(r_+^2-n^2)ge+4l^2n^2(r_+^4-4n^2r_+^2-n^4)g^2)\pi}{(l^2(n^2+r_+^2)e^2+4l^2n^2(n^2+r_+^2)g^2-(n^2+r_+^2)^2(l^2+3n^2+3r_+^2))} \nonumber\\
& \qquad +  \frac{(n^2+r_+^2)(l^2n^2-l^2r_+^2+3n^4-12n^2r_+^2-3r_+^4) \pi}
{(l^2 e^2+4l^2 n^2 g^2-(n^2+r_+^2) (l^2+3n^2+3r_+^2))}
\label{SNC-1} \\
N^{(1)}_{\cal N} &= -  \frac{(4 l^2 g^2 n^2 + l^2 e^2 - l^2 n^2 - l^2 r_+^2 - 3 n^4 - 6 n^2 r_+^2 - 3 r_+^4) n }{(n^2+r_+^2)r_+ l^2} \label{nut-1}\\
\psi^{(1)}_{\cal N} &= -\frac{n}{2} \frac{(4l^2g^2n^4+12l^2g^2n^2r_+^2+4l^2egn^2r_+-4l^2egr_+^3}{(4l^2g^2n^2+l^2e^2-l^2n^2-l^2r_+^2-3n^4-6n^2r_+^2-3r_+^4)(n^2+r_+^2)}
\nonumber\\
& \qquad - \frac{n}{2}\frac{l^2e^2n^2-l^2e^2r_+^2-l^2n^4-2l^2n^2r_+^2-l^2r_+^4-3n^6-3n^4r_+^2+3n^2r_+^4+3r_+^6)}{(4l^2g^2n^2+l^2e^2-l^2n^2-l^2r_+^2-3n^4-6n^2r_+^2-3r_+^4)(n^2+r_+^2)}
\label{psi-1}
 \end{align}
for the Noether charge entropy, the thermodynamic Noether NUT charge, and its conjugate potential.
It is straightforward to show that the first law
\be
dM = TdS^{(1)}_{{\cal N}} + \psi^{(1)}_{\cal N} dN_{\cal N}  + VdP +  \phi^{(1)}_e dQ_e  +  \phi^{(1)}_m dQ^{(+)}_m
\ee
and Smarr relation
\be\label{Smarr-1}
M = 2 T S^{(1)}_{{\cal N}}   - 2 PV +  \phi^{(1)}_e Q_e  +  \phi^{(1)}_m  Q^{(+)}_m
\ee
are both satisfied using \eqref{SNC-1}, \eqref{nut-1}, and \eqref{psi-1}. Note that the
$N_{\cal N}$ has no scaling dimension and so does not appear in \eqref{Smarr-1}.
\end{widetext}

The Noether charge entropy \eqref{SNC-1} depends on the parameters $(e,g)$ 
and  is not positive for all values of the parameters. This phenomenon has been seen before
in Taub-NUT AdS spacetimes \cite{ Mann:1999pc,Mann:1999bt} as well as in higher-curvature
gravity  \cite{Castro:2013pqa,Mir:2019ecg,Mir:2019rik}.  While negative entropy does not make  sense from a statistical mechanics viewpoint,  the LTN solution does not have any pathologies (beyond, perhaps, what we have noted  for the Misner string), and there is no obvious reason that these negative entropy solutions should be
rejected outright.   Furthermore, it is possible to shift the entropy by an arbitrary constant, either by adding
to the action a term proportional to the volume form of the induced metric on the horizon \cite{Clunan:2004tb}
or by including  an explicit Gauss-Bonnet term   \cite{Castro:2013pqa}.    This latter possibility has been studied in
some detail for the Euclidean Taub-NUT-AdS solution, where the introduction of the Gauss-Bonnet term was shown to
renormalize the Misner string contribution to the entropy   \cite{Ciambelli:2020qny}.

\begin{figure*}[t!]
    \centering
       \includegraphics[width=0.6\textwidth]{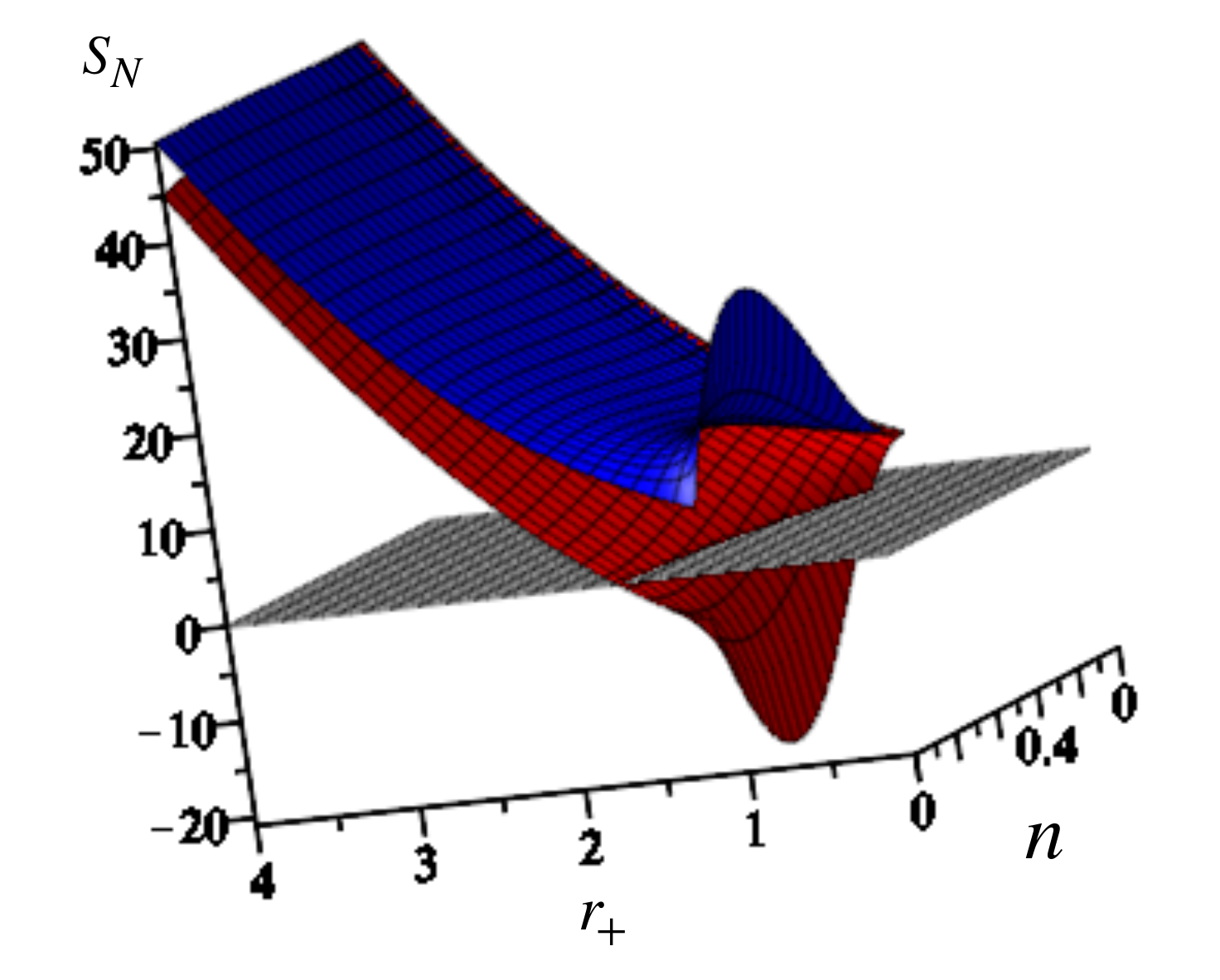}
    \caption{\textbf{Comparison of Noether charge Entropies for cases 1 and 2}~ The lower red sheet is case 1, the upper blue sheet is case 2, with $\ell = 30$, $g=0.5$, $e=1$.    The grey plane indicates zero.
}
\label{ent-compare} 
\end{figure*}

\section{Case 2: Horizon Electric Charge}

We now turn to  case 2, the   contrariwise situation for which 
the electric charge $Q_e^{(+)}\equiv q_e(r=r+)$, yielding
from \eqref{charge-e}  
\begin{equation}\label{case 2}
   Q_e^{(+)} =\frac{e(r_+^2-n^2)-4gr_+n^2}{r_+^2+n^2}
    \qquad Q_m=-2ng
\end{equation}
where now the magnetic charge is   given by \eqref{charges2}. 
The corresponding potentials are now given by :
\begin{align}\label{potentials-2}
&\phi^{(2)}_e=\frac{-2gn^2 + e r_+ }{r_+^2+n^2}\nonumber \\
&\phi^{(2)}_m= \frac{-n(2 g r_+ + e) }{r_+^2+n^2}  
\end{align}
and the thermodynamic NUT charges are
related to each other via
\begin{equation}
N^{(2)}_+ = N^{(1)}_+ - 16\pi n^2 \phi^{(2)}_e \phi^{(2)}_m
\end{equation}
by  electromagnetic duality.

From \eqref{XNpsi} and \eqref{SNS+}, we obtain
 \begin{widetext}
\begin{align}
S^{(2)}_{\cal N} &=\frac{\pi((6r^2+(4g^2-1)l^2)n^6+(24r^4- (16g^2+1)l^2r^2-8l^2egr+l^2e^2)n^4 -3n^8)}{(n^2+r^2)(3n^4-((4g^2-1)l^2-6r^2)n^2-l^2e^2+l^2r^2+3r^4)} \nonumber\\
&\qquad    +\frac{\pi((18r^6+(-4g^2+1)l^2r^4+8l^2egr^3+4l^2e^2r^2)n^2-l^2e^2r^4+l^2r^6+3r^8)}{(n^2+r^2)(3n^4-((4g^2-1)l^2-6r^2)n^2-l^2e^2+l^2r^2+3r^4)} 
\label{SNC-2} \\    
\psi^{(2)}_{\cal N} &= -\frac{n}{2} \frac{(-3 n^6+((4 g^2-1) l^2-3 r_+^2) n^4+(((-4 g^2-2) r_+^2-4 e g r_++e^2) l^2+3 r_+^4) n^2+(4 e g r_+^3+3 e^2 r_+^2-r_+^4) l^2+3 r_+^6)}{(-3 n^4+((4 g^2-1) l^2-6 r_+^2) n^2+(e^2-r_+^2) l^2-3 r_+^4) (n^2+r_+^2)}
  \label{psi-2}
 \end{align}
\end{widetext}
with the  thermodynamic Noether NUT charge still given by \eqref{nut-1}.

As before, the first law
\be
dM = TdS^{(2)}_{{\cal N}} + \psi^{(2)}_{\cal N} dN_{\cal N}  + VdP +  \phi^{(2)}_e dQ^{(+)}_e  +  \phi^{(2)}_m dQ_m
\ee
and Smarr relation
\be\label{Smarr-2}
M = 2 T S^{(2)}_{{\cal N}}   - 2 PV +  \phi^{(2)}_e Q^{(+)}_e  +  \phi^{(2)}_m  Q _m
\ee
are both satisfied using \eqref{SNC-2},  \eqref{psi-2}, and \eqref{nut-1}. The Noether
NUT charge $N_{\cal N}$ has no scaling dimension and so does not appear in \eqref{Smarr-2}.

We illustrate  in figure~\ref{ent-compare} a plot of the Noether charge entropy for cases 1 and 2. For sufficiently large $r_+$ they are similar, but notable distinctions appear at small $r_+$; these become more pronounced at larger $n$.  For nonzero $(e,g)$, the entropy diverges as a function of $(r_+,n)$ as 
$T\to 0$.  
 In both cases the Noether charge entropy \eqref{SNC-2}   is not
always positive for sufficiently small black holes due to the Misner string contributions as shown in
figure~\ref{ent-compare}.   Unlike the horizon magnetic case \eqref{SNC-1}, the entropy $S^{(2)}_{{\cal N}} $ is not a monotonic function of   $r_+$ -- as $r_+$ decreases the entropy decreases until it reaches a minimum, after which it increases further as $r_+$ decreases until the singularity is reached.

 \section{Case 3: Constrained Thermodynamics}

We now consider the situation in which the electromagnetic potential is constrained to vanish on the horizon.
This has generally been the approach taken for NUT-charged solutions in the Euclidean case
  \cite{Awad:2005ff,Dehghani:2006dk}, and   is equivalent to the  condition   $h(r_+)=0$, which is
  \begin{equation}\label{mag-e}
    g=-\frac{er_+}{r_+^2-n^2} 
\end{equation}
and using  \eqref{charges2} the
 asymptotic  electric charge and   electrostatic potential become
\begin{align}\label{case 3}
&Q=e \\
&\phi^{(3)}=-\frac{er_+-2 g n^2}{n^2+r_+^2} = -g
\nonumber 
\end{align}
from \eqref{charges2} and \eqref{phidiff}. The thermodynamic quantities \eqref{Mass}, \eqref{Temp} and \eqref{N1+} likewise must be changed to incorporate the constraint \eqref{mag-e}, and respectively become
\begin{equation}\label{Mass2}
 M^{(3)} = \frac{r^4_+ 6 n^2 r_+^2 - 3 n ^4  +(r_+^2 - n^2) l^2}{2 r_+ l^2} - \frac{e^2 (r_+^2+n^2)^2}{ 2(r_+^2 - n^2) r_+}
\end{equation}
\begin{equation}\label{Temp2}
 T^{(3)} = \frac{1}{4\pi r_+}\left(1+\frac{3(r_+^2+n^2)}{l^2}-\frac{e^2 (r_+^2+n^2)}{(r_+^2-n^2)^2}\right)
\end{equation}
and 
 \be 
N^{(3)}_+=-\frac{4\pi n^3}{r_+}\left(1+\frac{3(n^2-r_+^2)}{l^2}+ 
\frac{e^2 (3 r_+^2+n^2)}{(r_+^2-n^2)^2}\right)
\label{N3+}  
\ee 
and we see that $g$, $M$, and $T$ all become singular as $r_+ \to |n|$. However this occurs in an unphysical region where $T<0$.

The first law \eqref{flaw+} is no longer of full cohomogeneity, and reads
\be\label{flaw+2}
dM = T^{(3)}dS_+ + \psi_+ dN^{(3)}_+ + VdP +  \phi^{(3)}_e dQ_e   
\ee
and the Smarr relation
\be
M = 2 T^{(3)} S_+ + 2 \psi_+  N^{(3)}_+ - 2 VdP +  \phi^{(3)}_e dQ_e   
\ee
is likewise reduced.
\begin{figure*}[t!]
    \centering
      \includegraphics[width=0.6\textwidth]{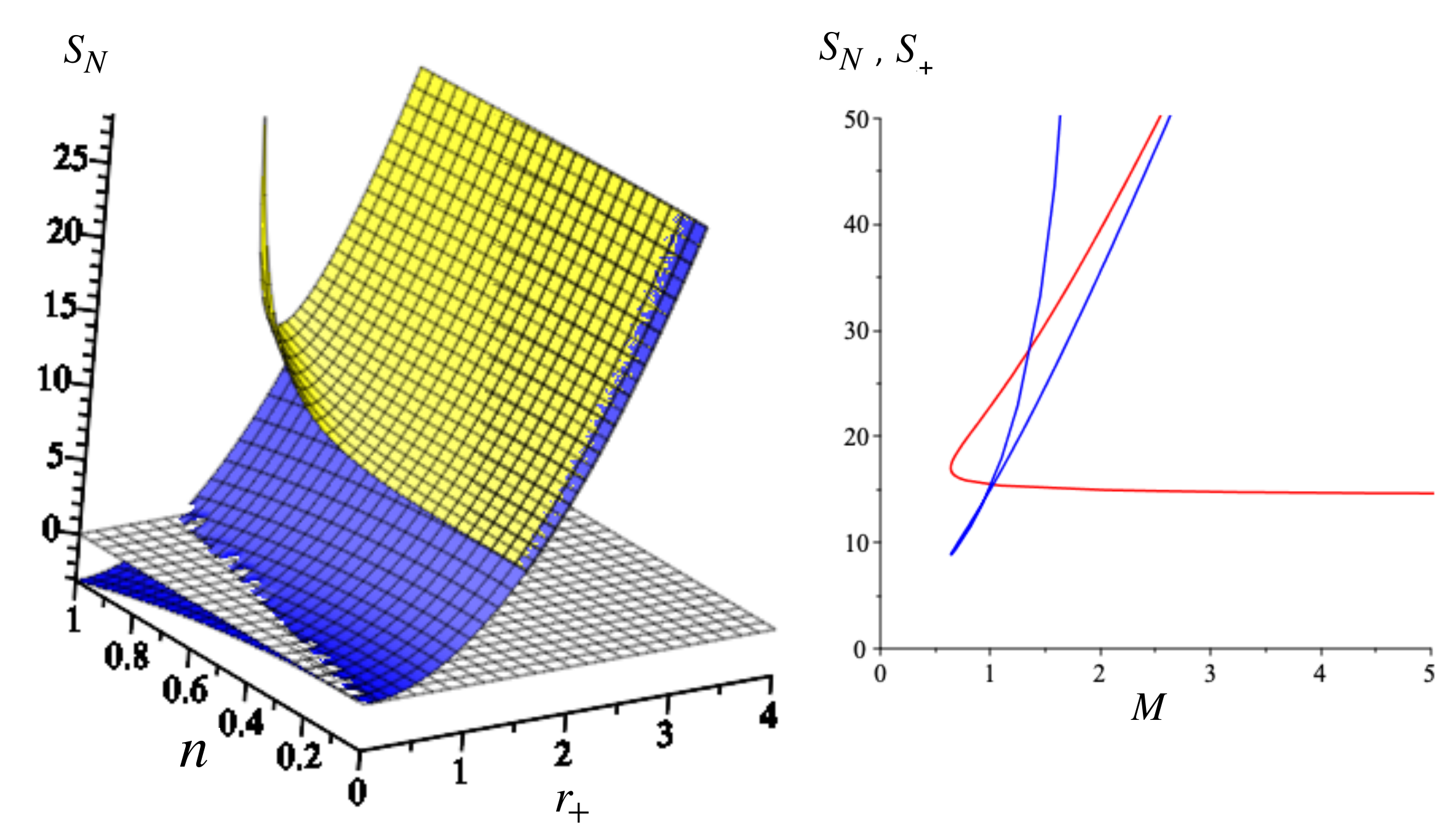}
    \caption{\textbf{Noether Charge Entropy for Case 3}~ Left: Two  plots of the Noether charge entropy $S_{\cal N}$  in case 3, for $\ell=12$, as a function of $(n,r_+)$.  The lower (blue) plot is    for $e=0$ and the upper (yellow) one for $e=0.6$.  The  grey plane indicates zero. Right: A plot of $S_{\cal N}$  (blue) and $S_{+}$ (red)
as a function of $M$ for $\ell = 7$, $e=.13$, and $n=1.5$.
   }
 \label{ent-fig-case3}
\end{figure*}

Once again, using \eqref{XNpsi} and \eqref{SNS+}, we obtain
 \begin{widetext}
\begin{align}
S^{(3)}_{\cal N} &=
 \frac{e^2 l^2 \pi ( r_+^4-n^4 - 4 n^2 r_+^2)  }{e^2 l^2 (n^2 + r_+^2) - (n^2 - r_+^2)^2 (l^2 + 3 (n^2 + r_+^2))}    -\frac{\pi (r_+^2-n^2)^2  (3 (r_+^4 + 4 n^2 r_+^2 - n^4)- l^2 (r_+^2-n^2) )}{e^2 l^2 (n^2 + r_+^2) - (n^2 - r_+^2)^2 (l^2 + 3 (n^2 + r_+^2))}
 \label{SNC-3} \\    
 N^{(3)}_{\cal N} &=   -n\frac{-l^2 (r_+^2+n^2) e^2+ (r_+^2 - n^2)^2 (l^2 + 3 n^2 +3 r_+^2) }{r_+ (r_+^2 - n^2)^2 l^2}
 \label{nut-3} \\
 \psi^{(3)}_{\cal N} &= -\frac{n}{2}  \frac{l^2 (n^2+3 r_+^2) e^2 -(r_+^2-n^2)^2 (l^2+3 n^2-3 r_+^2)}{e^2 l^2 (n^2 + r_+^2) - (n^2 - r_+^2)^2 (l^2 + 3 (n^2 + r_+^2))}
  \label{psi-3}
 \end{align}
\end{widetext}
where we note that $N^{(3)}_{\cal N}$ in \eqref{nut-3} is obtained from \eqref{nut-1} upon
inserting the constraint \eqref{mag-e}.  As before, the Noether charge entropy $S^{(3)}_{\cal N}$ 
in \eqref{SNC-3} depends on the charge parameter $e$  and
is not always positive.  This quantity is the same as  that obtained from the Euclidean section  \cite{Johnson:2014pwa} upon employing the Wick rotation $n\rightarrow in$ , $e\rightarrow ie$, and $g\rightarrow ig$.

We now have a first law of reduced cohomogeneity
\be
dM = T^{(3)} dS^{(3)}_{{\cal N}} + \psi^{(3)}_{\cal N} dN^{(3)} _{\cal N}  + VdP +  \phi^{(3)}_e dQ_e   
\ee
and   Smarr relation
\be\label{Smarr-3}
M = 2 T^{(3)}  S^{(3)}_{{\cal N}}   - 2 PV +  \phi^{(3)}_e Q_e   
\ee
that are straightforwardly shown to  both be satisfied using \eqref{Temp2}, \eqref{SNC-3},  \eqref{psi-3}, and \eqref{nut-3}. Once again, the Noether
NUT charge $N_{\cal N}$ has no scaling dimension and so does not appear in \eqref{Smarr-3}.

In this case we encounter a new phenomenon: neither the Noether charge entropy $S^{(3)}_{\cal N}$ nor
the horizon entropy $S_+$ is a single-valued function of $M$.  For $r_+ > n$, black holes of sufficiently small mass can
exist in either a high-entropy state or a low entropy state depending on the values of the other parameters.
This is because the mass $M$ is no longer a monotonically increasing function of the horizon radius -- for sufficiently small values of $M$, there are two allowed values (small and large) of $r_+$, yielding this behaviour.
If we admit solutions with $r_+ < n$, then a 3rd branch of solutions appears, in which 
  $S^{(3)}_{\cal N} < 0$ and  $S_+ > 0$ but smaller than the values in figure~\ref{ent-fig-case3}, each
an increasing function of $M$, with $S_+$ approaching its asymptotic value from below. 

\begin{widetext}
\phantom
{This table needs to be replaced with the correct one from the WORD file.}
\begin{table}[htp] 
\begin{center}
\caption{Free Energy and Phase Behaviour for Various Ensembles at fixed pressure $P$\\  
\phantom{$G^{(I)} = M - T S^{(I)}_{\cal N}$} $G^{(I)} = M - T S^{(I)}_{\cal N}$  $\qquad$  $G_\psi^{(I)} = M - T S^{(I)}_{\cal N}- \psi^{(I)}_{{\cal N}} N^{(I)}_{{\cal N}}$
}
\begin{tabular}{|c|c|c|c|c|c|c|}
\hline
Fixed Quantities & (a) $Q_e$, $Q_m$, $N^{(I)}_{{\cal N}}$ & (b) $Q_e$, $Q_m$,  $\psi^{(I)}_{\cal N}$   &(c)  $\phi_e$, $Q_m$, $N^{(I)}_{{\cal N}}$   &(d) $\phi_e$, $Q_m$, $\psi^{(I)}_{\cal N}$    &
(e) $\phi_m$, $Q_e$, $N^{(I)}_{{\cal N}}$   &(f) $\phi_m$, $Q_e$, $\psi^{(I)}_{\cal N}$ 
 \\ [0.5ex] 
 \hline\hline
Horizon Magnetic  & $F = G^{(1)}$ & $F=G_\psi^{(1)} $ & $F = G^{(1)} - \phi_e Q_e$ & $ F = G_\psi^{(1)}  - \phi_e Q_e$ & $ F = G^{(1)} - \phi_m Q_m$ & 
$F = G_\psi^{(1)} - \phi_m Q_m$  \\
(Case I) &   &     &  &  &    &   \\
\hline
Horizon Electric & $F = G^{(2)}$ & $F=G_\psi^{(2)} $ & $F = G^{(2)} - \phi_e Q_e$ & $ F = G_\psi^{(2)}  - \phi_e Q_e$ & $ F = G^{(2)} - \phi_m Q_m$ & 
$F = G_\psi^{(2)} - \phi_m Q_m$   \\
(Case II)  &    &   &     &     &    &      \\
\hline
Constrained  & $F = G^{(3)}$ & $F=G_\psi^{(3)} $  & $F = G^{(3)} - \phi_e Q_e$ & $ F = G_\psi^{(3)}  -\phi_e Q_e$ & &    \\
(Case III) &   &   &   &   &  &   \\
\hline
\end{tabular}
\end{center}
\end{table}

\section{Phase Behaviour and Thermodynamic Ensembles}

As noted in the introduction, we consider six distinct thermodynamic ensembles for each case. This is summarized in table I.  One feature common to all cases is the notion of a threshold pressure that governs the behaviour of the temperature.  Solving either of case I or case II for $(e,g)$ in terms of $(Q_e,  Q_m)$ yields
 \begin{align}
T  & = \frac{3r_+^6 + (l^2 - 3n^2)r_+^4 - ((Q_e^2 + Q_m^2 + 2 n^2) l^2 + 3 n^4)r_+^2 - 4 l^2 Q_e Q_m n r_+  +  3 n^6( 1-P_t/P)}{4\pi l^2 r_+ (r_+^2-n^2)^2 }
\label{Tcase1}
\end{align}
where
\be
P_t = \frac{Q_e^2 + Q_m^2 - n^2}{8\pi n^4} \; .
\label{Ptcase1}
\ee
is a threshold pressure.   For case I,  $Q_e$ and $Q_m$ in \eqref{Tcase1}  and \eqref {Ptcase1}   are given by  \eqref{case 1}, whereas for case II
they are given by \eqref{case 2}, but with $Q_m$ replaced by $-Q_m$.  For case III the threshold pressure is given
by replacing $Q_m \to 0$ in \eqref {Ptcase1}.
\end{widetext}
\begin{figure*}[t!]
	\centering
	\includegraphics[scale = 0.3]{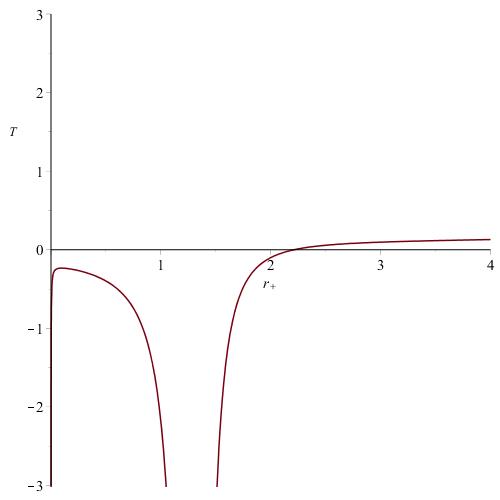} 
	\includegraphics[scale = 0.3]{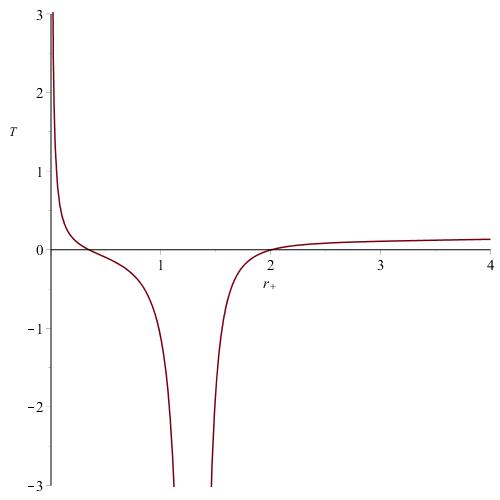} 
 \includegraphics[scale = 0.3]{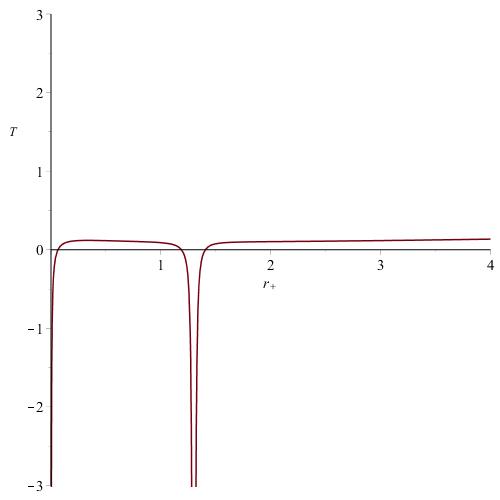} \\
  \includegraphics[scale = 0.3]{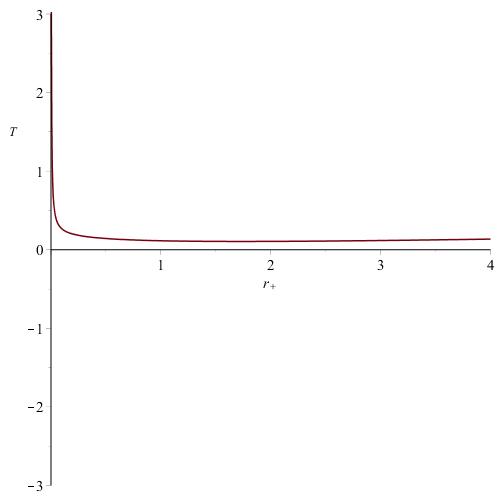} 
   \includegraphics[scale = 0.3]{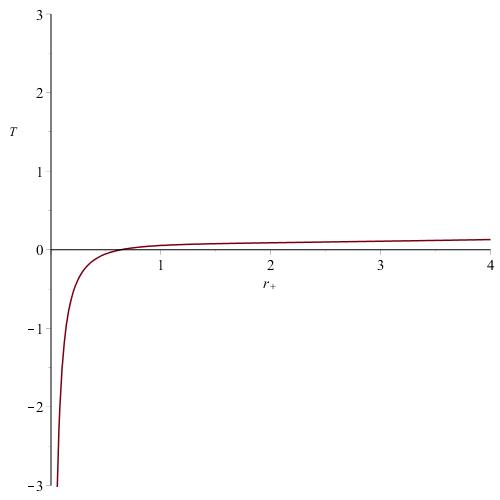}
	\caption{ Possible behaviours of the temperature for various values of the parameters, with $\ell =3$. 
The upper three figures all have singularities at $r_+ = n$.	 The upper left figure has $P_t > P$, with  $Q_e=1$, $Q_m=1.3$  and  $n=1.3$. There is only one extremal black hole.   The upper middle figure has $P_t < P$, with $Q_e=1$, $Q_m=0.7$ and $n=1.3$. There is both a large and a small extremal black hole. The upper right figure has $P_t > P$, with $Q_e=-1.1$, $Q_m=1.3$ and $n=1.3$. There is now a large extremal black hole and two smaller extremal black holes.  The lower left figure has $P_t < P$, with $Q_m=-Q_e=-1.1$,  and $n=1.3$; in this case there are no extremal black holes.  The lower right figure has $P_t > P$, with $Q_m=-Q_e=-1.4$,  and $n=1.3$; in this case there is one extremal black hole, but no singularity in $T$ for $r_+ > 0$.   	}
	\label{figtemp}
\end{figure*}

 \begin{figure*}[t!]
	\centering
	\includegraphics[scale = 0.4]{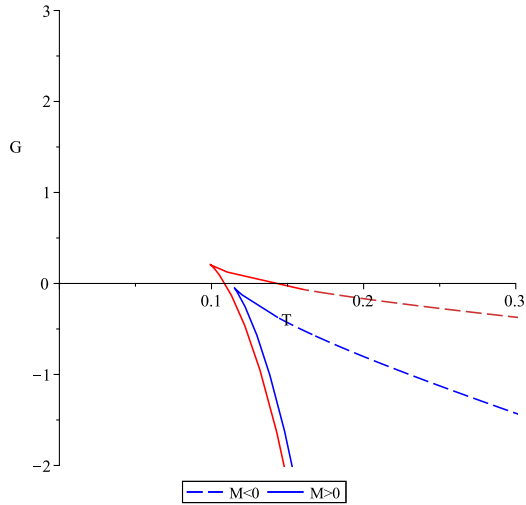}  
	\caption{{\bf Cusp Structure for vanishing charge} Setting $e=g=0$,
we observe a cusp structure for all values of $\ell$ and $n$.	Here we set  $\ell =3$;   $n=1.3$  and $n=0.7$ are the blue and red curves respectively, with the dash corresponding to $M<0$ solutions.  
	}
	\label{case3Q0}
\end{figure*}

 \begin{figure*}[t!]
	\centering
	\includegraphics[scale = 0.5]{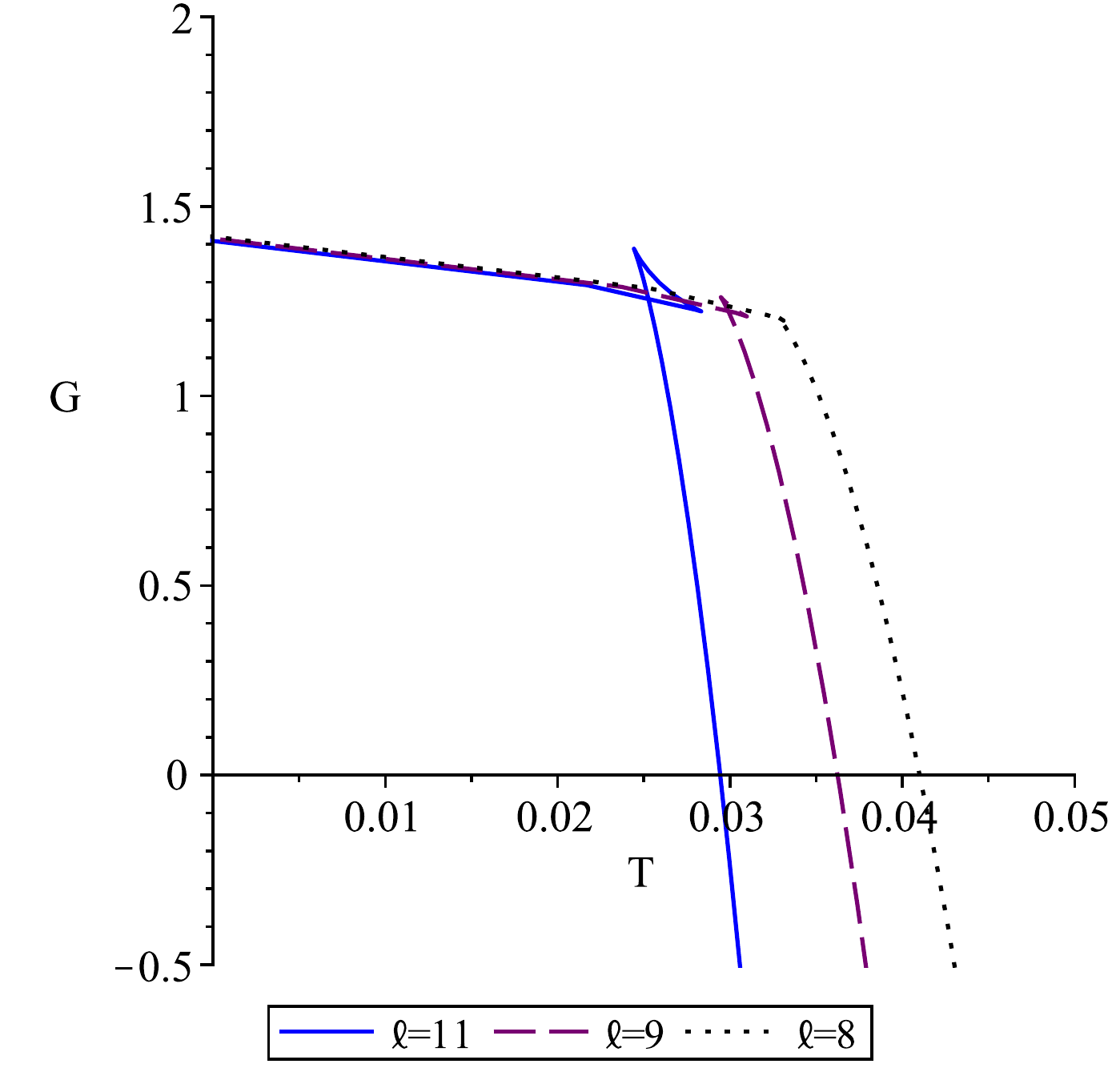} 
	\includegraphics[scale = 0.5]{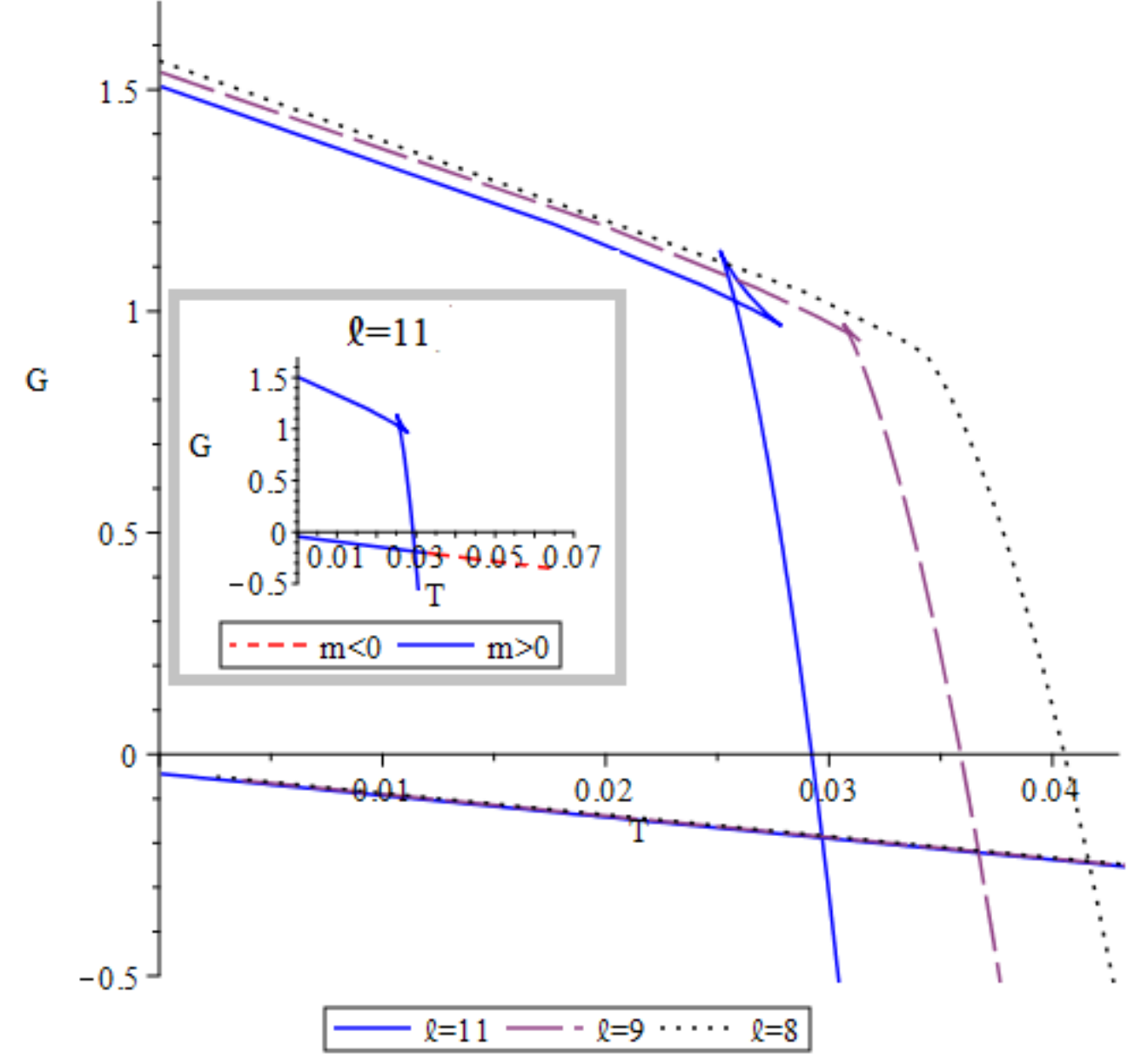} 
	\caption{{\bf Swallowtails below and above threshold pressure} {Left: Setting $n=1$, $Q_e=1.4$ and $Q_m=-1.3$,
	we have $P_t =0.10544 >P$ for all values of $P$ (or $\ell$)  shown. This situation exhibits the low-pressure swallowtail and high-pressure single-phase behaviour that takes place for $n=0$ Reissner-Nordstrom AdS black holes.  At the intersection of the swallowtail, there is a first-order phase transition from a large black hole to a small one as the temperature decreases.
Right: Setting $n=1.2$, $Q_e=1$ and $Q_m=0.1$,  we have $P_t<0$, ensuring $P>P_t$ for all values of $P$ shown (or $\ell$). 
There are now two branches. For one branch  $r_+ > n$; the curves exhibit the same qualitative behaviour as the diagram at the left. Along the other branch  $r_+ < n$; for all values
	of $P$ these branches are nearly indistinguishable.  This branch has a lower free energy than the $r_+ > n$ branch  at low temperatures, and so there will be a first order phase transition where these two branches intersect. However this branch 
will also have negative mass for sufficiently small $r_+$, illustrated in the inset.  These structures occur in cases Ia and IIa.
	}}
	\label{case1-a}
\end{figure*}  

\begin{figure*}[t!]
\centering
\includegraphics[scale = 0.3]{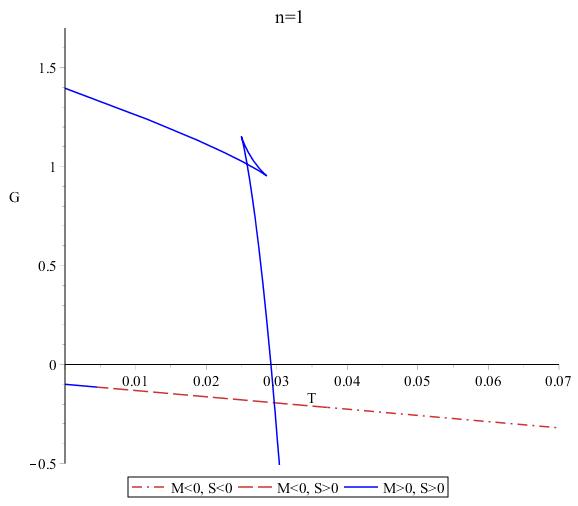} 
\includegraphics[scale = 0.3]{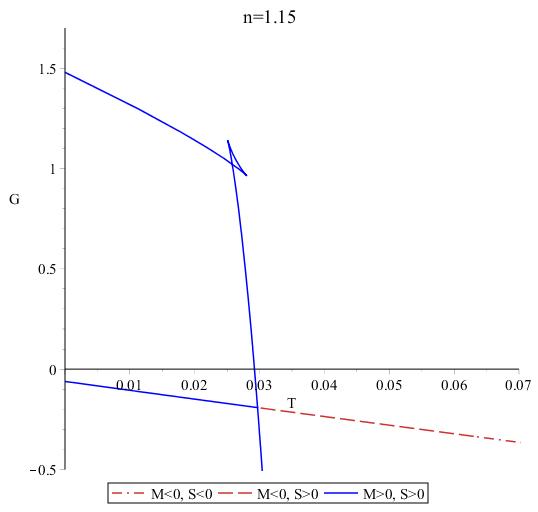} 
\includegraphics[scale = 0.3]{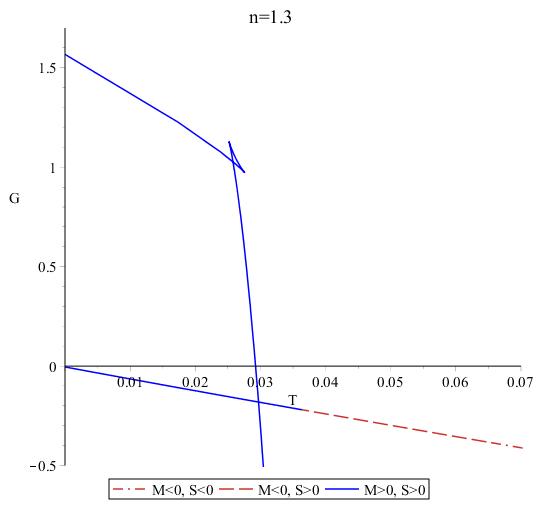} 
\caption{{\bf Interrupted Swallowtail} {For $\ell=11$, $Q_e=1$ and a range of values of $n$, we observe an interrupted swallowtail, where the usual large/small transition is unstable, and instead 
a large/tiny first order phase transition takes place as temperature decreases. 
Blue solid lines correspond to $M>0$, whereas red dashed lines correspond to $M<0$. 
 For large enough $n$
this transition will be for $M>0$, shown in the right diagram. As $n$ decreases, it reaches a threshold value ($n=1.15$, shown in
the middle diagram) below which the transition is from a large black hole with $M>0$ to a
the tiny black hole with $M<0$, shown in the left diagram. Throughout, the entropy $S^{(I)}_{\cal N} > 0$ at the transition; 
the unstable branch at large $T$ eventually has $S^{(I)}_{\cal N} < 0$, shown by the red dot-dash line. These structures occur in cases Ia and IIa.
}}
\label{case1-a2-interrupt}
\end{figure*}
\begin{figure*}[t!]
	\centering
	\includegraphics[scale = 0.3]{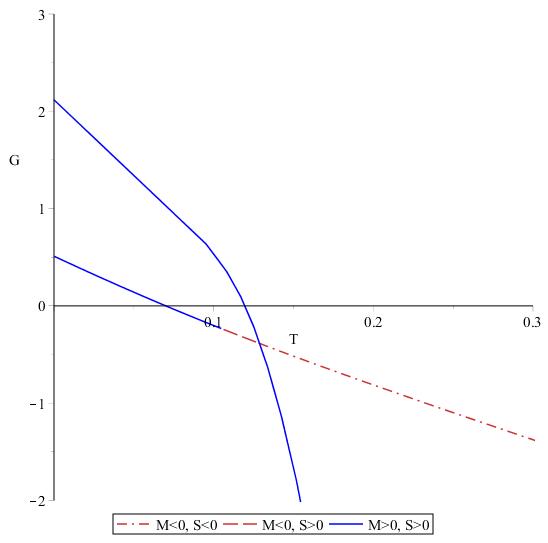} 
	\includegraphics[scale = 0.3]{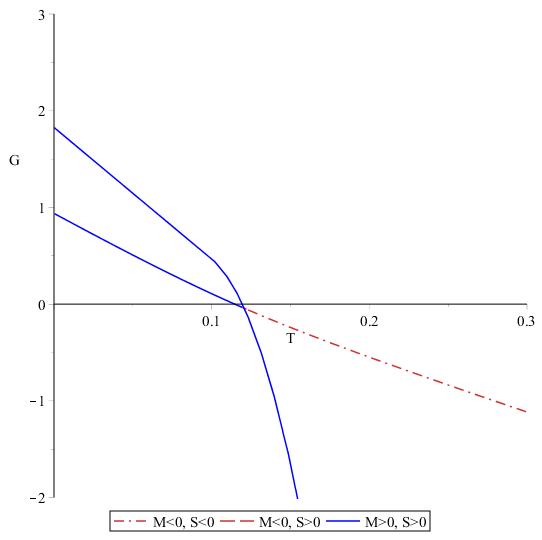} 
	\includegraphics[scale = 0.3]{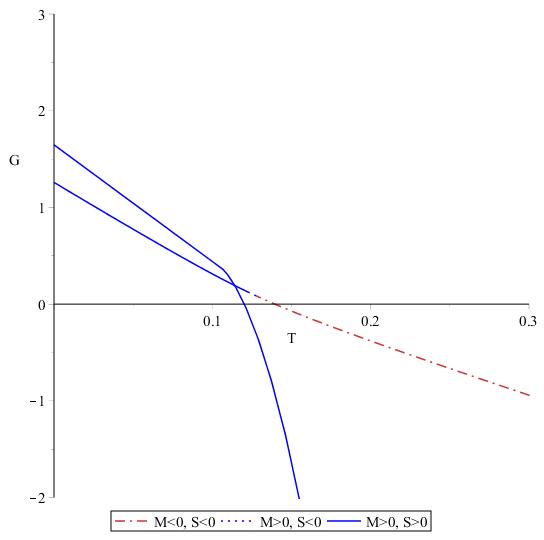} 
	\caption{ {\bf Large/Tiny transitions} For sufficiently high pressure (here $\ell =3$), 
	the swallow tail is absent, and only a large/tiny transition takes place.  Setting   $Q_e=1$ and $n=1.3$,
	we observe a situation analogous to that in figure~\ref{case1-a2-interrupt}, but with increasingly negative
	magnetic charge governing whether or not $M<0$ for the tiny phase.  Proceeding left to right, the left diagram has    $Q_m=-0.2$, the middle one $Q_m=-0.55$ and the  right one $Q_m=-0.8$.   Blue solid lines correspond to $M>0$ and $S>0$, red dashed lines   to $M<0$, and $S>0$, and red dot-dash lines to $M<0$, and $S<0$.
	}
	\label{largetiny}
\end{figure*}

\begin{figure*}[t!]
	\centering
	\includegraphics[scale = 0.4]{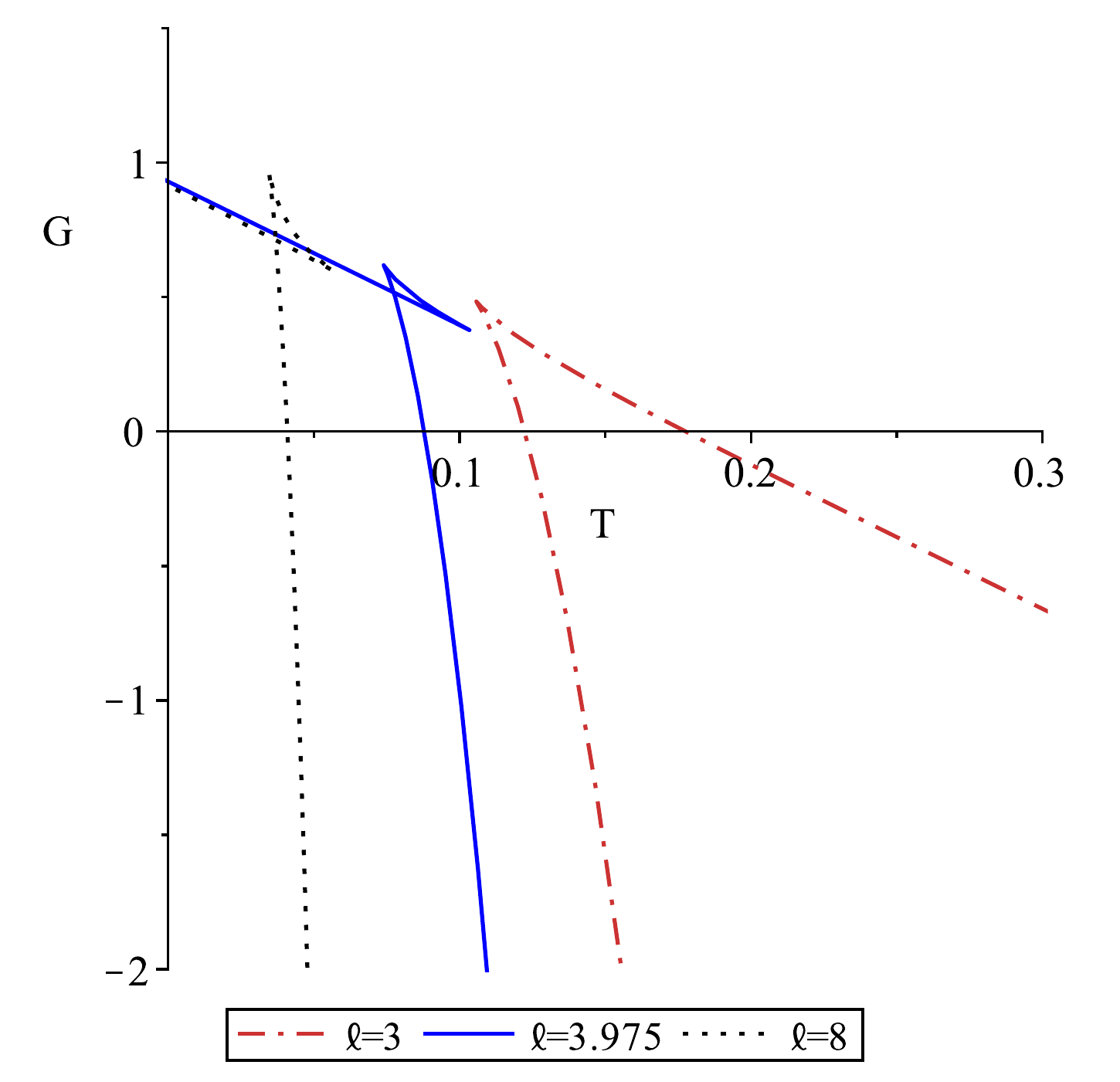} 
	\caption{{\bf Breaking Swallowtail} {Here we set $n=1.3$ and $Q_e=-Q_m=1.1$. The larger two values of $\ell$ correspond to  $P< P_t$, and exhibit the familiar swallowtail behaviour. However for $P>P_t$ ($\ell=3$) the swallowtail is replaced by a cusp. No second order critical behaviour is observed. This behaviour can occur in case Ia.
}}
	\label{case1-a-eqchg}
\end{figure*}

\begin{figure*}[t!]
	\centering
	\includegraphics[scale = 0.5]{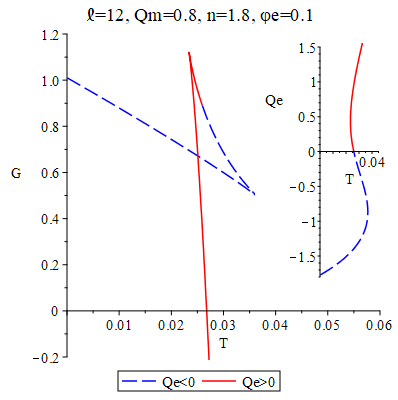} 
	\caption{{\bf Charge-Changing Phase Transition}  { Plotting free energy as a function of temperature for $Q_m=0.8$, $n=1.8$, $\phi_e=0.1$, and $\ell=12$,  we observe a first order large/small phase transition, where the electric charge changes sign. The inset shows the behaviour of the charge as a function of temperature. This behaviour is typical of case Ic, and for its magnetic counterpart, case Ie.		}
	}
	\label{case1-c1}
\end{figure*} 
\begin{figure*}[t!]
	\centering
	\includegraphics[scale = 0.5]{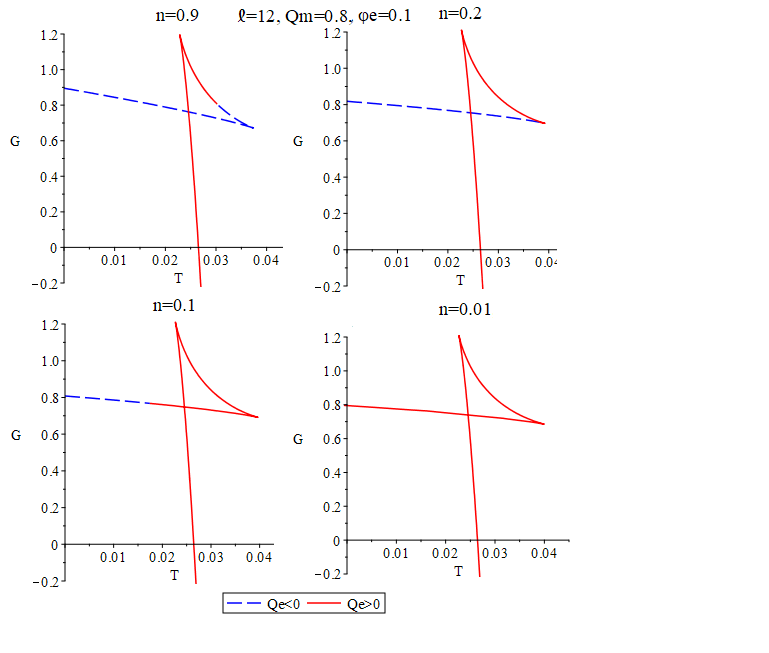} 
	\caption{{\bf NUT charge dependence of Charge-changing Transitions} {We see that as $n$ increases in size, the small branch becomes negatively charged. The charge transition point eventually moves to the unstable part of the swallowtail, at which point charge-changing transitions become possible.	 Throughout  all these diagrams, $M>0$. This behaviour 
	can occur in cases Ic and Ie.} 
	}
	\label{case1-c2}
\end{figure*}

We shall employ the notation $G^{(I)} = M - T S^{(I)}_{\cal N}$, for which $N^{(I)}_{{\cal N}}$ 
is fixed and 
 $G_\psi^{(I)} = M - T S^{(I)}_{\cal N}- \psi^{(I)}_{{\cal N}} N^{(I)}_{{\cal N}}$ for which $\psi^{(I)}_{{\cal N}}$
 is fixed, the index $I = 1,2,3$ denoting the respective horizon magnetic, horizon electric, and constrained
 cases.   Note that using \eqref{SNS+}, we have 
\begin{align}\label{G-S}
G^{(I)} &= M - T S^{(I)}_{\cal N} = M - T S_+  - \psi_+ N_+ \\
G_\psi^{(I)} &= M - T S^{(I)}_{\cal N}- \psi^{(I)}_{{\cal N}} N^{(I)}_{{\cal N}} = M - T S_+   {- 2\psi_+ N_+} 
\label{G-S2}
\end{align}
indicating  for each case that fixed $N^{(I)}_{{\cal N}}$ corresponds to fixed $\psi_+$ (or fixed $n$).  

We shall now discuss the various phase transitions that can occur.

\subsection{Fixed $N^{(I)}_{{\cal N}}$}
\label{FixedN}

As stated above, this case corresponds to ensembles in which one either regards the entropy as being $S^{(I)}_{\cal N}$, or
in which $n$ (or $\psi_+$) is fixed and the entropy is $S_+$.  The phase structures obtained are equivalent in either case.
This corresponds to columns (a), (c), and (e) in the table.

The form of \eqref{Tcase1} indicates that there can be as many as 3 extremal black holes depending on the magnitudes relative signs  of $Q_e$, $Q_m$,  and $P_t/P$.  The various possibilities are illustrated in figure~\ref{figtemp}.
We see that if $P_t > P$ then $T\to\infty$ as $r_+\to 0$, whereas if $P_t < P$ then $T\to -\infty$ as $r_+\to 0$. The rule of signs can be used to infer the remaining behaviour.  If $P_t > P$ and $Q_e$ and $Q_m$ have the same sign then $T$ has  two positive roots, corresponding to the upper middle diagram in figure~\ref{figtemp}. However if $P_t < P$ then $T$ has either one root if $Q_e$ and $Q_m$ have the same sign (shown in the upper left diagram) or three roots if they have opposite sign 
(shown in the upper left diagram).
Two interesting special cases occur if $Q_e=-Q_m$.  The singularity at $r_+ = n$ is removed, and the temperature either has one root if $P_t > P$ (shown in the lower right diagram in figure~\ref{figtemp}) or no roots if  $P_t < P$ (shown in the lower left diagram).

For case III, the upper right and lower right diagrams in figure~\ref{figtemp} are not possible. For vanishing charge, only the lower left diagram is possible, as there are no extremal black holes. For any nonzero charge there will either be one or two extremal black holes, corresponding to the upper left and upper middle diagrams respectively. 
For all possibilities in case III we  observe behaviour similar to that of cases I and II where these remaining diagrams in figure~\ref{figtemp} are applicable, and so we shall not illustrate this case in what follows. 

 Since physical solutions must have positive temperature, we will obtain various branches of possible physical solutions for
each of the various possibilities. This will have interesting implications for the free energy and phase behaviour of the LTN black hole as we shall see.  Note that  while $T$
 is singular at $r_+ = n$ (as are $M$ and $N_{\cal N}$) if  $Q_e\neq -Q_m$, this  takes place in an unphysical region where $T<0$.  

\subsubsection{Vanishing Charge}

For zero magnetic and electric charge,  the free energy diagram corresponds to that of a cusp, shown in
figure~\ref{case3Q0} for two different values of $n$.  This cusp may be above or below the $G=0$ axis in the free-energy diagram, with larger values of $n$ moving the cusp to smaller values of
$G$.  If $n=0$ the intersection of the cusp signifies a Hawking-Page transition to thermal AdS (radiation). 
But if $n\neq 0$ this cannot take place in the $G^{(I)}$ ensembles, since these correspond to fixed $n$, whereas thermal AdS
has $n=0$.   
 
\subsubsection{Interrupted Swallowtails}

For cases Ia and IIa, we have a phenomenon that we refer to as the `interrupted swallowtail',   previously observed  \cite{Ballon:2019uha}  for Lorentzian NUT-charged AdS black holes
in which $S_+$ is taken to be the entropy.    We illustrate this in figure~\ref{case1-a}.    

If $P_t  > P $,   the classic swallowtail structure observed for $n=0$ Reissner-Nordstrom AdS black holes \cite{Kubiznak:2012wp} takes place, as shown in the left diagram in
figure~\ref{case1-a}.  For low pressures there is a first order large/small phase transition as the temperature decreases; for high pressures there is only a single phase.    However if $P  >  P_t $  then  an additional  new branch
appears  for small $r_+ < n $ for which the free energy is negative.  In this case the first order phase transition at the swallowtail interaction will not take place. Instead there will be a first order  transition at the intersection of the $r_+ >n$ branch with the $r_+ < n $ branch, as shown in the right diagram in figure~\ref{case1-a}.  The would-be swallowtail transition is `interrupted' by the 
lower branch transition -- essentially the large/small transition becomes a large/tiny transition.

 There is a caveat to this, however.  The mass on this lower branch is not always positive -- as temperature increases the mass can become negative.  If negative mass solutions are not ruled out as unphysical, then the first order phase transition above will take place.  However 
if they are ruled out, then this will not take place and the usual large/small swallowtail transition takes place.  
This is shown in the inset in the left diagram in figure~\ref{case1-a}.  In this particular case, the pressure is such that the negative mass tiny solutions have higher free energy than the large solutions, and so a large/tiny transition will take place. However 
as the pressure increases, the negative mass part moves toward lower temperatures on the $r_+<n$ branch and the large/tiny transition will not take place if negative mass solutions are ruled out \cite{Ballon:2019uha}.  

We illustrate this in figure~\ref{case1-a2-interrupt}.
Depending on where this occurs, as temperature decreases there will either be a zeroth order large/tiny transition (if the negative mass solutions set in at a temperature larger than the swallowtail intersection) or a first order large/small transition (at the swallowtail intersection) followed by a zeroth order small/tiny transition (if the negative mass solutions set in at a temperature smaller than the swallowtail intersection).   
For sufficiently high pressure the swallowtail is absent, and only a large/tiny first order transition takes place, shown in
figure~\ref{largetiny}.

\subsubsection{Breaking Swallowtails}

If we set $Q_e=-Q_m$, then we have behaviour associated with the lower two diagrams in figure~\ref{figtemp}; we illustrate this behaviour in figure~\ref{case1-a-eqchg}. Here  the swallowtail `breaks' in a manner similar to the snapping behaviour seen 
for accelerating black holes \cite{Abbasvandi:2018vsh}. 
For $P< P_t$ an extremal black hole exists, and the free energy diagram exhibits  swallowtail behaviour with the familiar
first order large/small transition as temperature decreases \cite{Kubiznak:2012wp}.  However as the pressure increases so that $P> P_t$, instead of a critical point being reached, the swallowtail becomes a cusp.  No phase transition takes place; instead, as temperature decreases the large black hole becomes smaller until it attains a minimal size below which 
it is  thermodynamically unstable; in  figure~\ref{case1-a-eqchg}, threshold at which $P = P_t$ occurs at $\ell = 3.975$.

\subsubsection{Charge-changing Phase Transitions}

Further interesting behaviour occurs if we consider fixed electromagentic potentials. In this case it is possible to get
first order transitions from large positively charged black holes to small negatively charged ones (and vice-versa, depending on the parameter choices made.

To be specific, we illustrate in figure~\ref{case1-c1} a typical situation for  fixed $\phi_e$ and $Q_m$, corresponding to case
Ic.  As the temperature decreases,  there is a first-order large/small phase transition where the sign of $Q_e$ changes.
Essentially the black hole discharges all its positive charge and picks up negative charge from the fixed potential.
If we fix  $Q_e$ and $\phi_m$ (Case Ie) we observe similar behaviour, but with the magnetic charge $Q_m$ changing sign.

 In figure~\ref{case1-c2} we illustrate how this charge-changing behaviour changes as the fixed value of $n$ is varied. For sufficiently small $n$ no charge-changing behaviour takes place.  As $n$ increases, the small branch develops negative charge, which grows along the small branch until eventually the charge-changing first order transition takes place.

\subsubsection{Inverted Cusps}

We also observe a structure we refer to as an inverted cusp.  This can occur  if $P<P_t$ and the charges are sufficiently large and of opposite sign.  In this case the temperature as a function of $r_+$ is given by the upper right diagram in figure~\ref{figtemp}.  There are now three extremal black holes, with the smaller two yielding an inverted cusp structure in the free energy.
\begin{figure*}[t!]
	\centering
	\includegraphics[scale = 0.45]{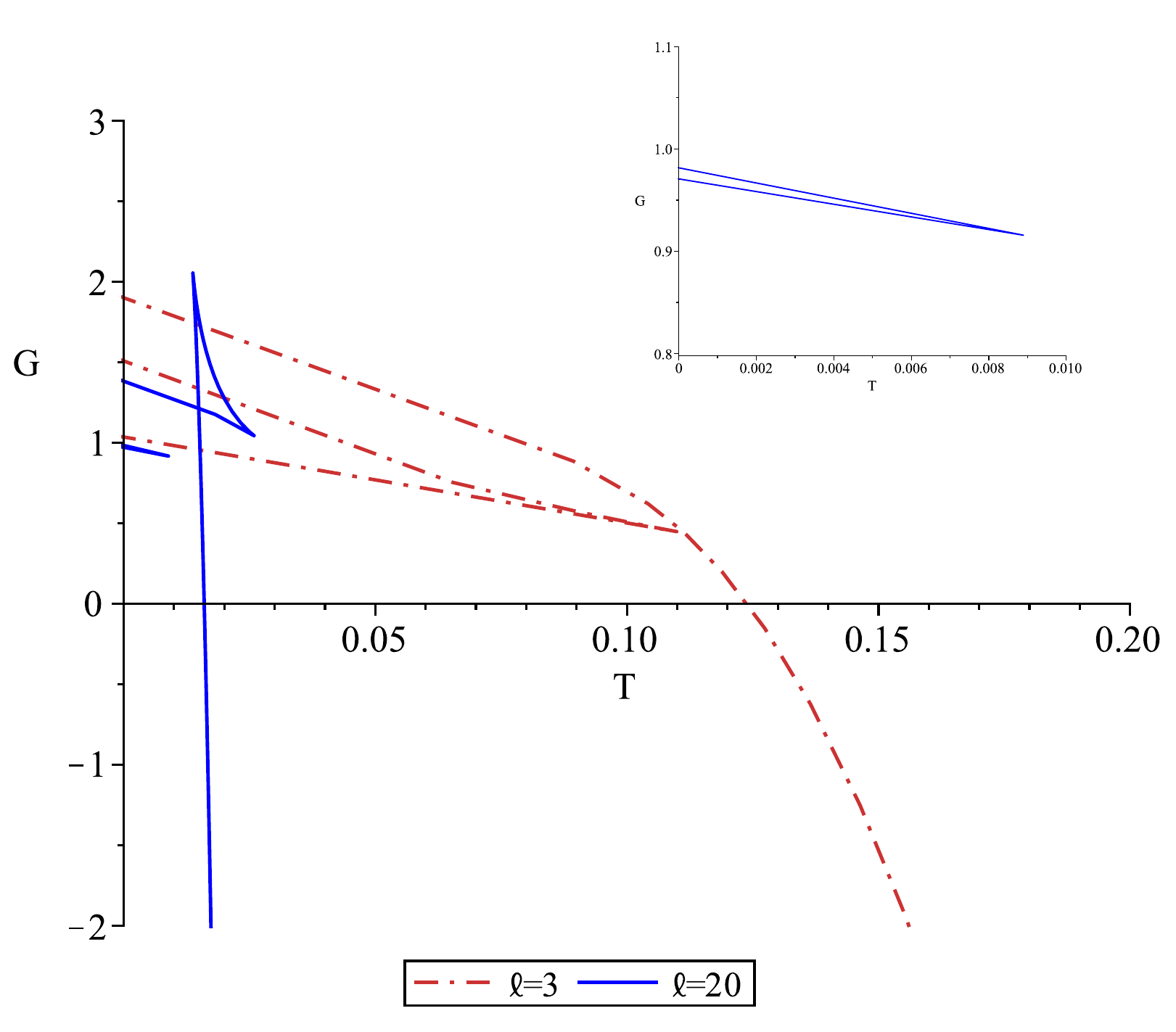} 
	\includegraphics[scale = 0.45]{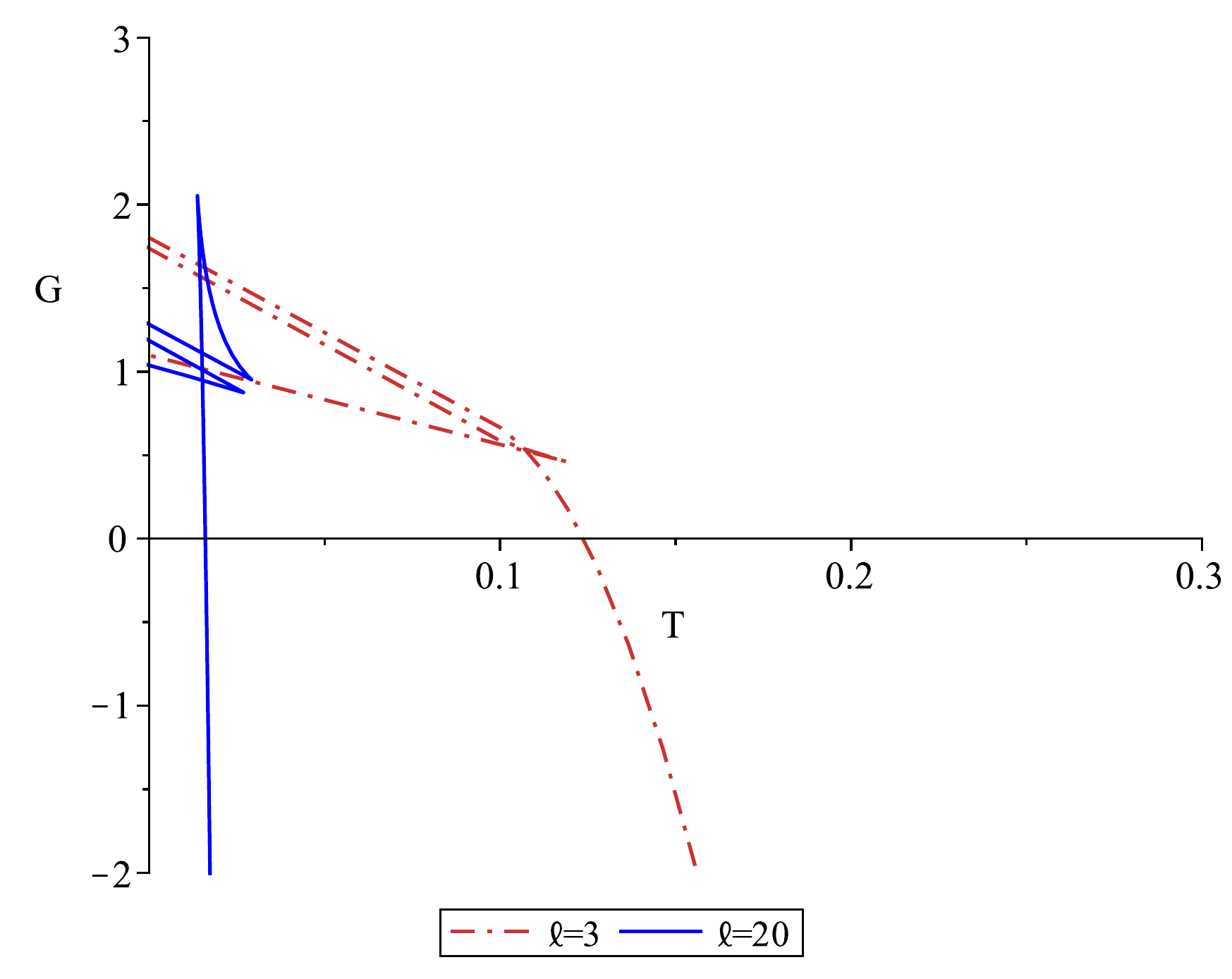} 
	\caption{{\bf Inverted Cusp} {For $P<P_t$ we observe inverted cusp structures when there are three possible extremal black holes. Blue solid curves correspond to $\ell=20$ (low pressure) and red dot-dash curves to $\ell=3$ (high pressure). Left: We set $n=1.3$, $Q_e = -1.39$, and $Q_m =0.969$. The inverted cusps do not intersect the other curves corresponding to larger black holes.  The inset provides a close up of the cusp for $\ell=20$. Right: We set $n=1.3$, $Q_e = -1.30$, and $Q_m =1.10$. The inverted cusps  intersect the other curves corresponding to larger black holes. These situations can occur in cases Ia and Ib.} 
	}
	\label{invcusp1}
\end{figure*} 

This is illustrated in figure~\ref{invcusp1}.  There is a branch of large black holes ($r_+ > n$) that exhibits the standard swallowtail structure at low pressures ($\ell=20$ in the figure), which vanishes at a critical point at a sufficiently high pressure, beyond which the curve is smooth ($\ell=3$ in the figure).   However there is now branch of tiny black holes ($r_+ < n$) that has the form of an inverted cusp.  This cusp may or may not intersect the large branch, depending on the relative size of the magnetic charge compared to the electric one.  

If the magnetic charge is sufficiently small, the cusps do not intersect, as shown in the left diagram of figure~\ref{invcusp1}. 
In this case, at low pressure there will be a standard first order large/small phase transition as the temperature decreases;
above and below the transition $r_+>n$.  This will be followed by a zeroth order   transition, in which the small
$r_+>n$ black hole becomes a tiny $r_+ < n$ black hole. The lower branch of the cusp corresponds to the smallest of the tiny black holes and is thermodynamically stable.  At high pressures there is no swallowtail, but the  zeroth order   transition will still
take place. 

If the magnetic charge is sufficiently large, the cusps  intersect, as shown in the right diagram of figure~\ref{invcusp1}. 
As temperature decreases there is no longer a first order large/small transition at low pressure. Instead, for all pressures, there is a first order phase transition  in which the large $r_+>n$ black hole becomes a tiny $r_+ < n$ black hole.

\begin{figure*}[t!]
	\centering
	\includegraphics[scale = 0.45]{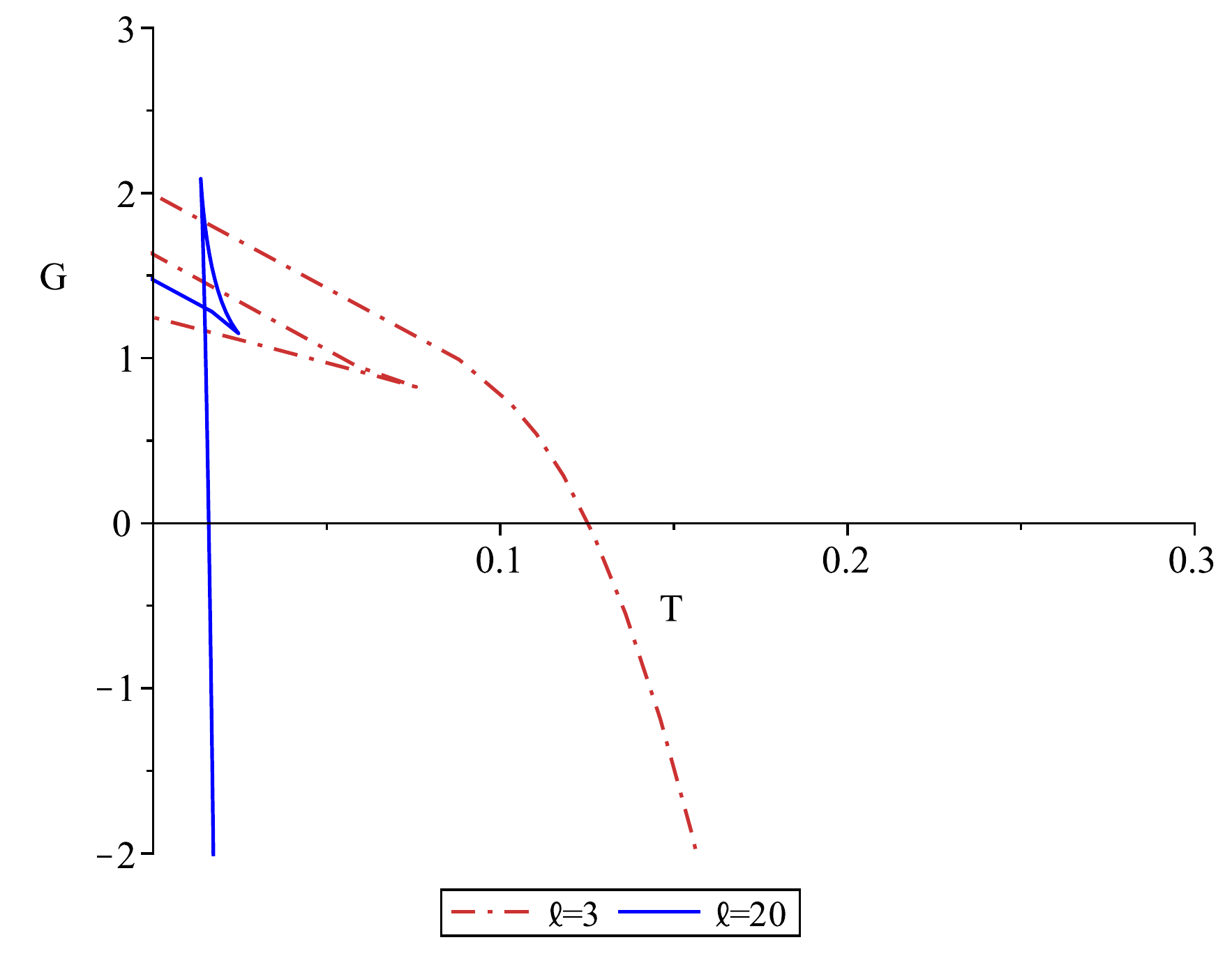} 
	\includegraphics[scale = 0.45]{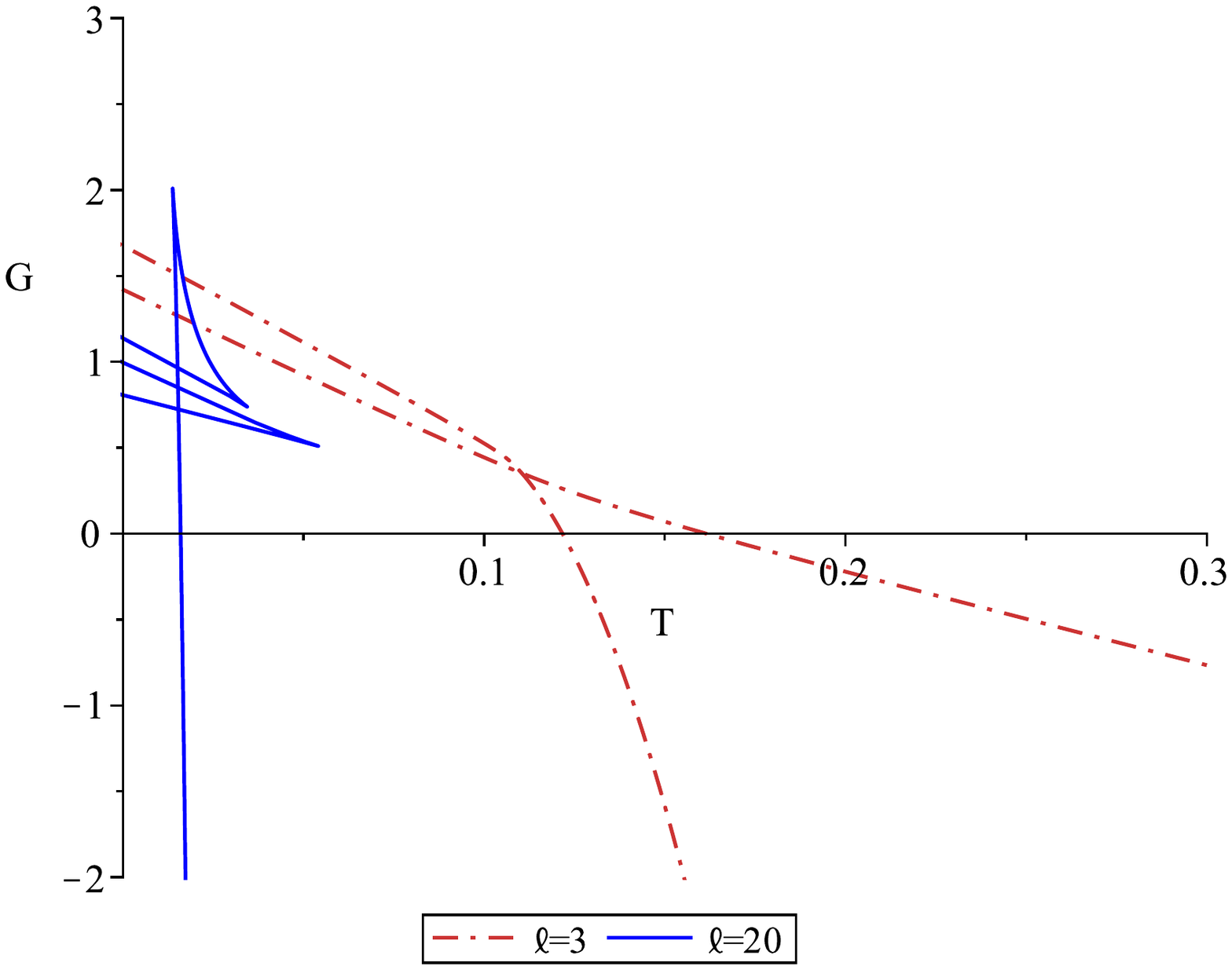} 
	\caption{{\bf Pressure Dependent Inverted Cusps} { Blue solid curves correspond to $\ell=20$ (low pressure) and red dot-dash curves to $\ell=3$ (high pressure). Left: We set $n=1.3$, $Q_e = -1.5$, and $Q_m =1.1$.  In this case there is only one extremal black hole at low pressure, but three extremal black holes at high pressure.  Right: We set $n=1.3$, $Q_e = -1.1$, and $Q_m =0.964$.  In this case there are three extremal black holes at low pressure, but only one extremal black hole at high pressure.} 
	}
	\label{invcusp2}
\end{figure*} 
It is also possible to have situations in which three extremal black holes are present at low pressure but not high pressure, and vice-versa,  as shown in figure~\ref{invcusp2}.  In this case the phase behaviour will be a combination of the previous cases.
In the left diagram, as temperature decreases we observe a large/small first order transition where $r_+>n$ at low pressure and a zeroth order large/tiny transition at high pressure.  In the right diagram, for both pressures we have only a first order 
large/tiny transition as temperature decreases.

These structures change as the pressure changes.  In the left diagram in figure~\ref{invcusp2}, as pressure decreases, the inverted cusp recedes, eventually vanishing. A swallowtail then develops as pressure further increases. The right diagram in figure~\ref{invcusp2} has a somewhat more complicated behaviour. As pressure decreases, the lower $r_+<n$ branch shifts a bit and then suddenly snaps to an inverted cusp that intersects the large branch curve. As pressure further decreases, a swallowtail forms on the large branch above the inverted cusp, giving rise to the low pressure swallowtail plus inverted cusp structure we see in the diagram.

Finally, it is possible for the inverted cusp to just intersect the large black hole branch. In this case there will be a form
of triple point, where the large, tiny, and unstable tiny phases coexist.

\subsection{Fixed $ \psi^{(I)}_{{\cal N}}$}
\label{FixedPsi}

For   fixed $ \psi^{(I)}_{{\cal N}}$ we obtain distinct ensembles from those considered previously  \cite{Ballon:2019uha}, corresponding to columns (b), (d), and (e) in the table.
The entropy is now interpreted as $S^{(I)}_{\cal N}$, and only $G_\psi^{(I)}  = M - T S^{(I)}_{\cal N}- \psi^{(I)}_{{\cal N}} N^{(I)}_{{\cal N}} $ has a sensible interpretation.

\begin{figure*}[t!]
    \centering
    \includegraphics[scale = 0.35]{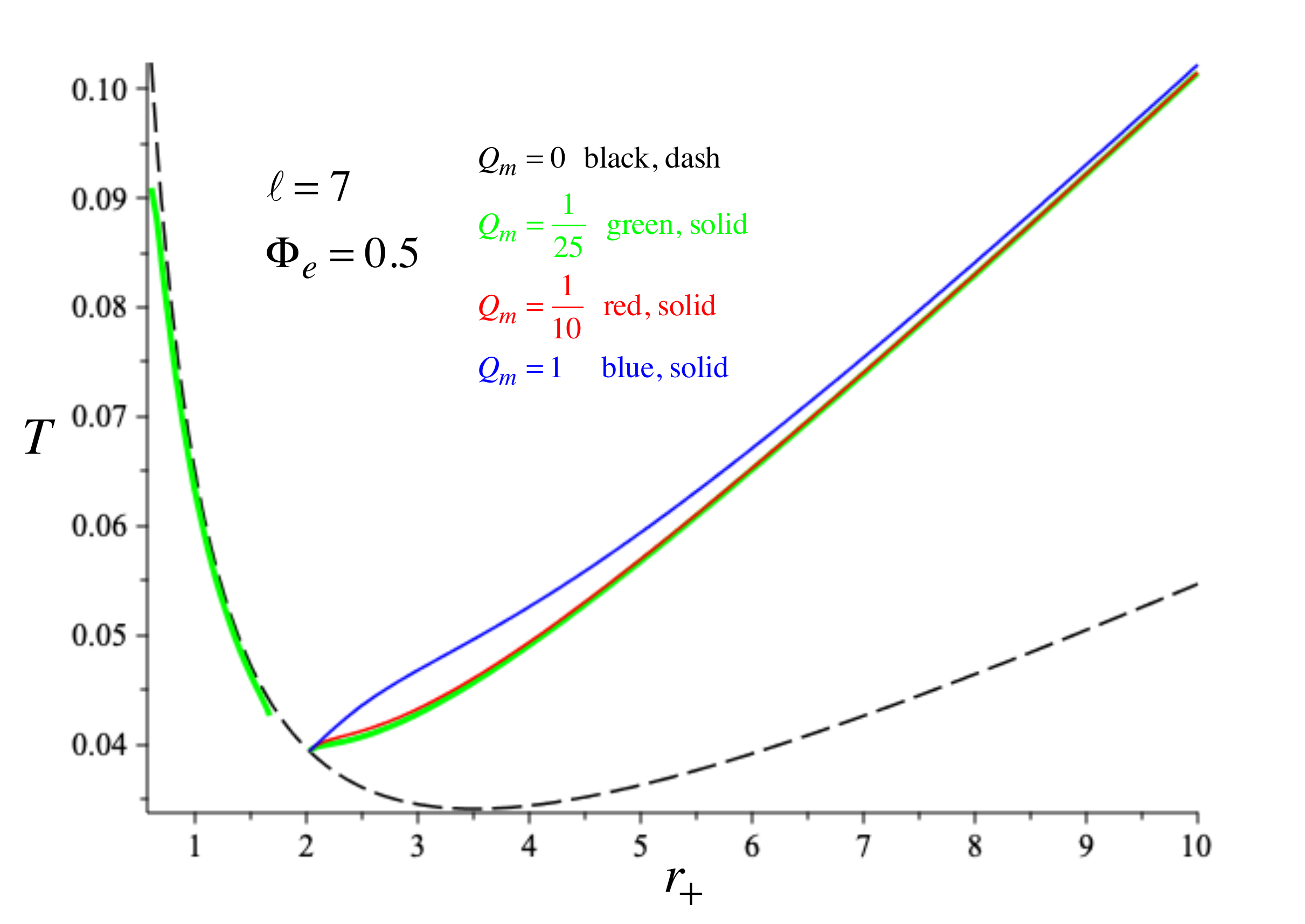} 
    \caption{{\bf Temperature vs. Horizon size for nonzero magnetic charge at fixed electric potential}  This plot of  $T$ vs. $r_+$ at $ \psi^{(I)}_{{\cal N}} = 0$ exhibits a discontinuity
    in temperature for various values of $Q_m$ relative to the $Q_m=0$ case. For sufficiently large $Q_m$ black hole solutions do not exist at small $r_+$.   
   Here $\ell = 7$ and $\Phi_e = 0.5$.
     }
    \label{case1Tvsrp}
\end{figure*}

\begin{figure*}[t!]
    \centering
    \includegraphics[scale = 0.35]{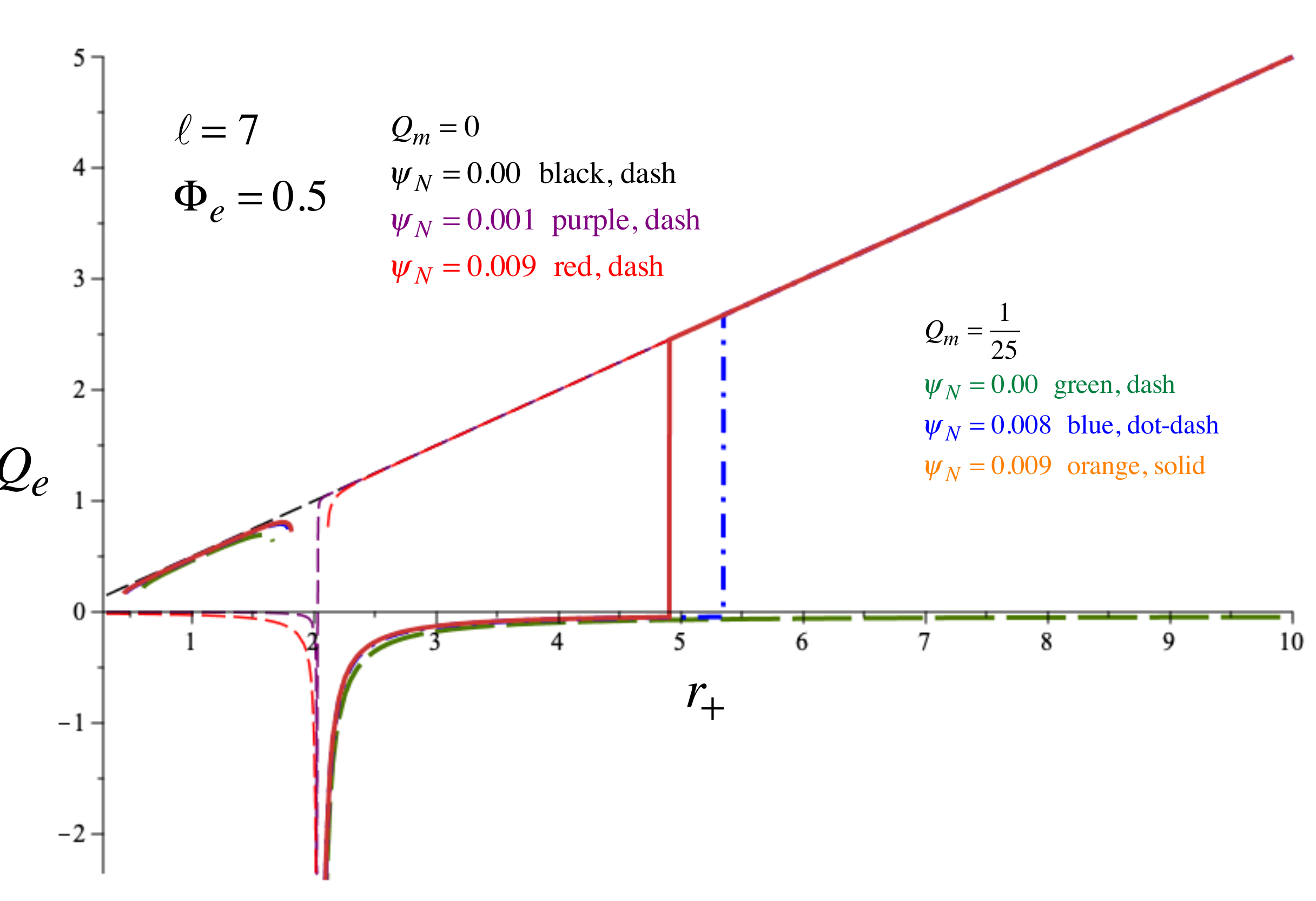} 
    \caption{{\bf Electric charge vs. Horizon size at fixed electric potential}  This plot of  $Q_e$ vs. $r_+$ at $ \psi^{(I)}_{{\cal N}} = 0$ exhibits a single discontinuity
    for vanishing magnetic charge $Q_m=0$, that widens as $ \psi^{(I)}_{{\cal N}} $ increases, with a positively charged black hole becoming negatively charged at sufficiently small $r_+$.  If  $Q_m \neq 0$
    a second discontinuity appears if  $ \psi^{(I)}_{{\cal N}} \neq 0$, and there is an intermediate range of $r_+$ where $Q_e < 0$. This region narrows as $ \psi^{(I)}_{{\cal N}}$ increases.
   Here $\ell = 7$ and $\Phi_e = 0.5$.
     }
    \label{case1Qevsrp}
\end{figure*}

Before proceeding to describe the additional qualitatively new phase behaviour we observe, we first note that discontinuities are present in both temperature and charge when plotted as functions of $r_+$. This is illustrated in figure~\ref{case1Tvsrp} for
the temperature and in figure~\ref{case1Qevsrp} for the electric charge, for various values of the magnetic charge. This behaviour undergirds the various phase behaviours for fixed $ \psi^{(I)}_{{\cal N}}$ that we now go on to describe.

 \subsubsection{Inverted Swallowtails}

\begin{figure*}[t!]
    \centering
    \includegraphics[scale = 0.5]{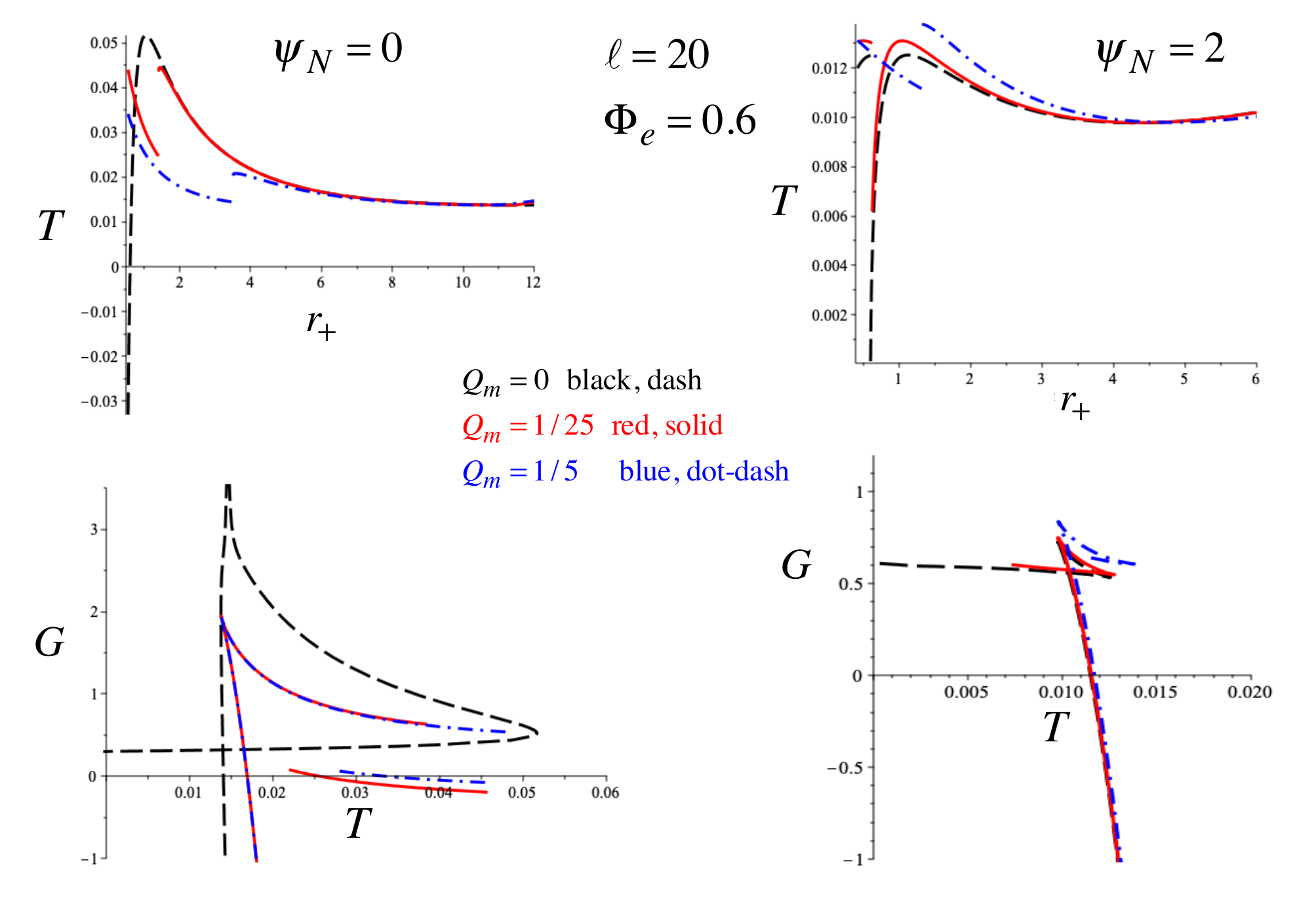} 
    \caption{{\bf Case 1(b) Fixed Electric and Magnetic Charges }~ The plots on the left are for temperature and free energy for fixed $\psi^{(I)}_{\cal N}=0$,
    and on the right for fixed $\psi^{(I)}_{\cal N} = 2$, and for the respective fixed horizon magnetic charges shown.  The fixed  electric charge $Q_e=0.6$,
    and $\ell = 20$.  The presence of magnetic charge induces a discontinuity in the free energy, but this is restored for sufficiently large $\psi^{(I)}_{\cal N}$.
  }
    \label{case-1b}
\end{figure*}

 Consider first fixed electric and magnetic charges with fixed $\psi^{(I)}_{\cal N}$, column (b) in the table.  If $\psi^{(I)}_{\cal N}=0$, the familiar swallowtail structure appears for vanishing magnetic charge, corresponding to the familiar first-order large/small black hole phase transition.  However  fractures appear for nonzero $Q_m$, as shown in figure~\ref{case-1b}. In this case no phase transition is present.   This swallowtail structure is restored for sufficiently large $\psi^{(I)}_{\cal N}$.
 
More generally this cases exhibits a double swallowtail structure, with one swallowtail inverted, as shown in figure~\ref{case1-b1}.  For large pressures only the lower inverted swallowtail is present, but for small enough pressure the upper cusp becomes a swallowtail. As  pressure decreases, the lower swallowtail structure shrinks and the upper one grows.   The formation of
the inverted swallowtail as a function of increasing magnetic charge is shown in figure~\ref{case1-b4}.

This situation exhibits interesting phase behaviour. At high temperatures we have a large black hole.  As temperature decreases, there will be a first order transition to a small black hole, on the lower part of the curve containing the inverted swallowtail.  These black holes may have $M<0$, depending on the choice of parameters.  As temperature further decreases, the small black hole grows in size, eventually undergoing a zeroth order phase transition to a larger black hole on the bottom of the inverted swallowtail.  This black hole may undergo a further zeroth order  transition to an even larger black hole 
if the pressure is sufficiently large (the $\ell=7$ curve in figure~\ref{case1-b1}), or else remain at some fixed value beyond which no black holes exist if the pressure is smaller (the $\ell=8$ curve in figure~\ref{case1-b1}).  If the pressure is small enough
for an upper swallowtail to be present, then there will be a zeroth order transition to a larger black hole on the lower branch of the upper swallowtail (the $\ell=9,10$ curves in figure~\ref{case1-b1}). As temperature further decreases, the hole shrinks in size a bit, terminating at some value of $r_+$ and $T$ below which no black holes can exist. This latter behaviour is more easily seen in figure~\ref{case1-b2}.

 {\tcb{
	 \begin{figure*}[t!]
		\centering
		\includegraphics[scale = 0.5]{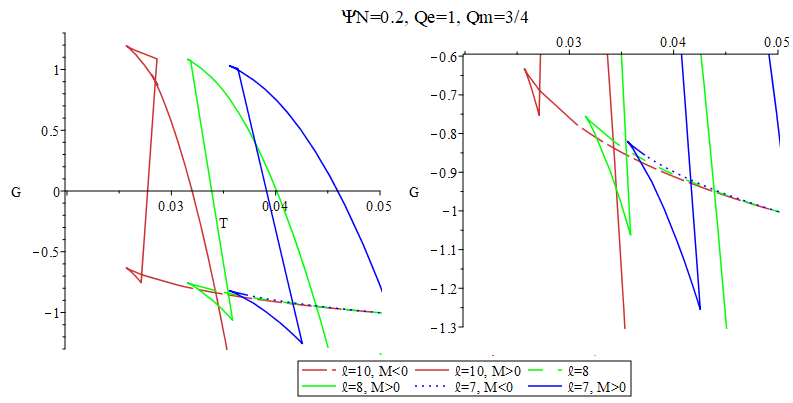} 
		\caption{{\bf Inverted Swallowtails}  These structures can appear for cases Ib and IIb; we illustrate them here for
		$Qe=1$ and $Q_m=0.75$ (case Ib).  At large pressures there is a lower swallowtail on what would otherwise be a cusp.  For sufficiently large $\ell$, the upper cusp is replaced with a swallowtail.  As $\ell$ increases (or as pressure decreases)
		the lower swallowtail  shrinks and the upper one grows.  Note that the lower parts of each curve have segments where
		$M<0$.  	The insets provide close-ups of the upper and lower parts of the curves.  
		}
		\label{case1-b1}
	\end{figure*} 
 \begin{figure*}
	\centering
	\includegraphics[scale = 0.5]{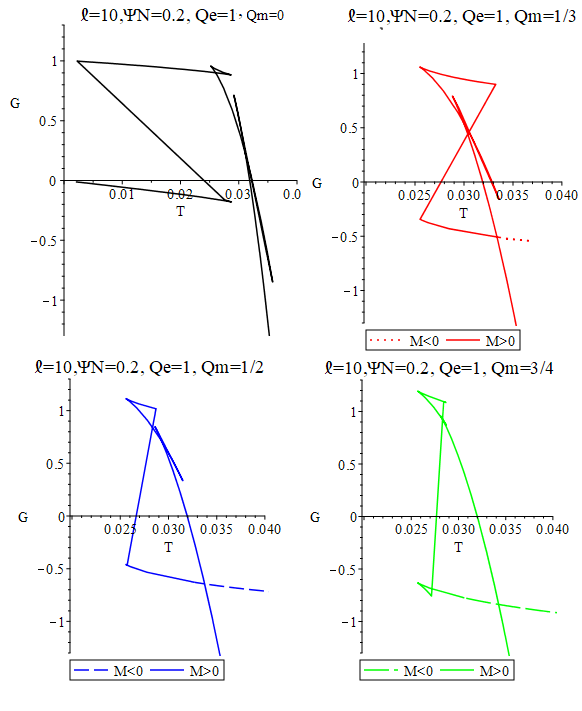} 
	\caption{{\bf Formation of Inverted Swallowtails} This sequence of diagrams shows how the inverted swallowtails form
	as the magnetic charge increases. Note the appearance of an additional narrow swallowtail on the unstable large black hole branch at the right for smaller values of $Q_m$.  	}
	\label{case1-b4}
\end{figure*} 
 \begin{figure*}[t!]
	\centering
	\includegraphics[scale = 0.6]{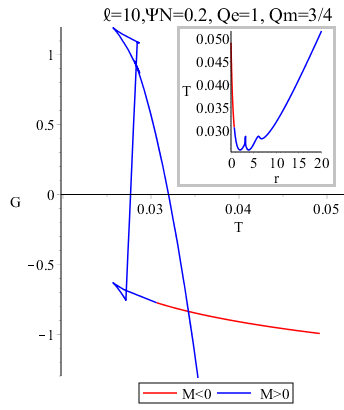} 
	\caption{{\bf Inverted Swallowtail Transitions}  This particular case undergoes three phase transitions as temperature decreases. The first is a large $M>0$ to small $M<0$ black hole, shown at the rightmost intersection.  As temperature decreases, the mass becomes positive.  Eventually there is a zeroth order transition to a slightly larger black hole, located at the rightmost cusp of the lower inverted swallowtail.  As temperature further decreases, the hole gets larger, eventually undergoing another zeroth order transition to an even larger black hole located on the bottom branch of the upper swallowtail. The temperature finally decreases to its minimal value, below which no black hole exist for this choice of parameters. The inset shows the behaviour of the temperature as a function of horizon size.  Note that there are no extremal black holes.
 }
	\label{case1-b2}
\end{figure*} 
}} 

\begin{figure*}[t!]
    \centering
    \includegraphics[scale = 0.45]{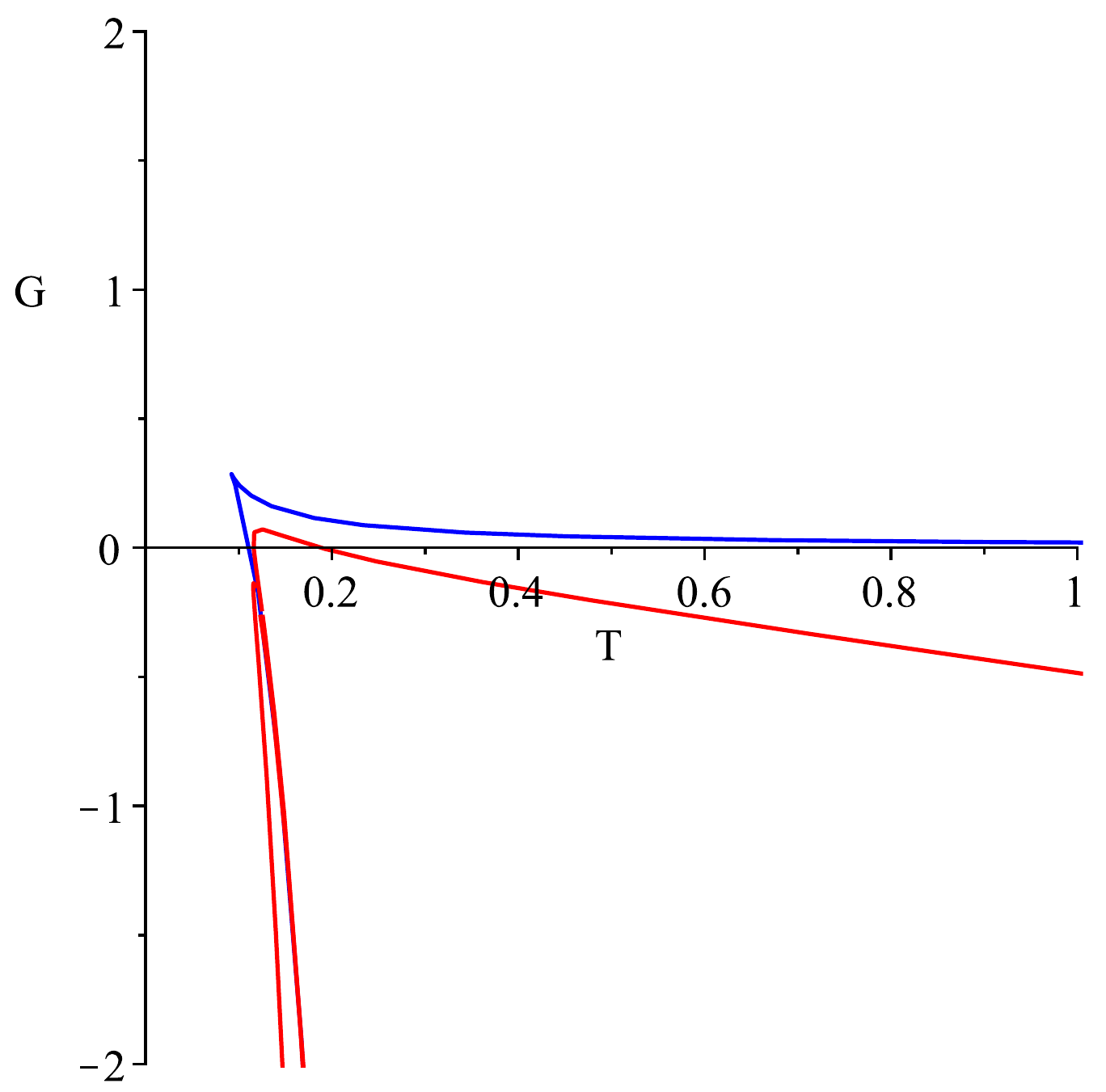} 
      \includegraphics[scale = 0.45]{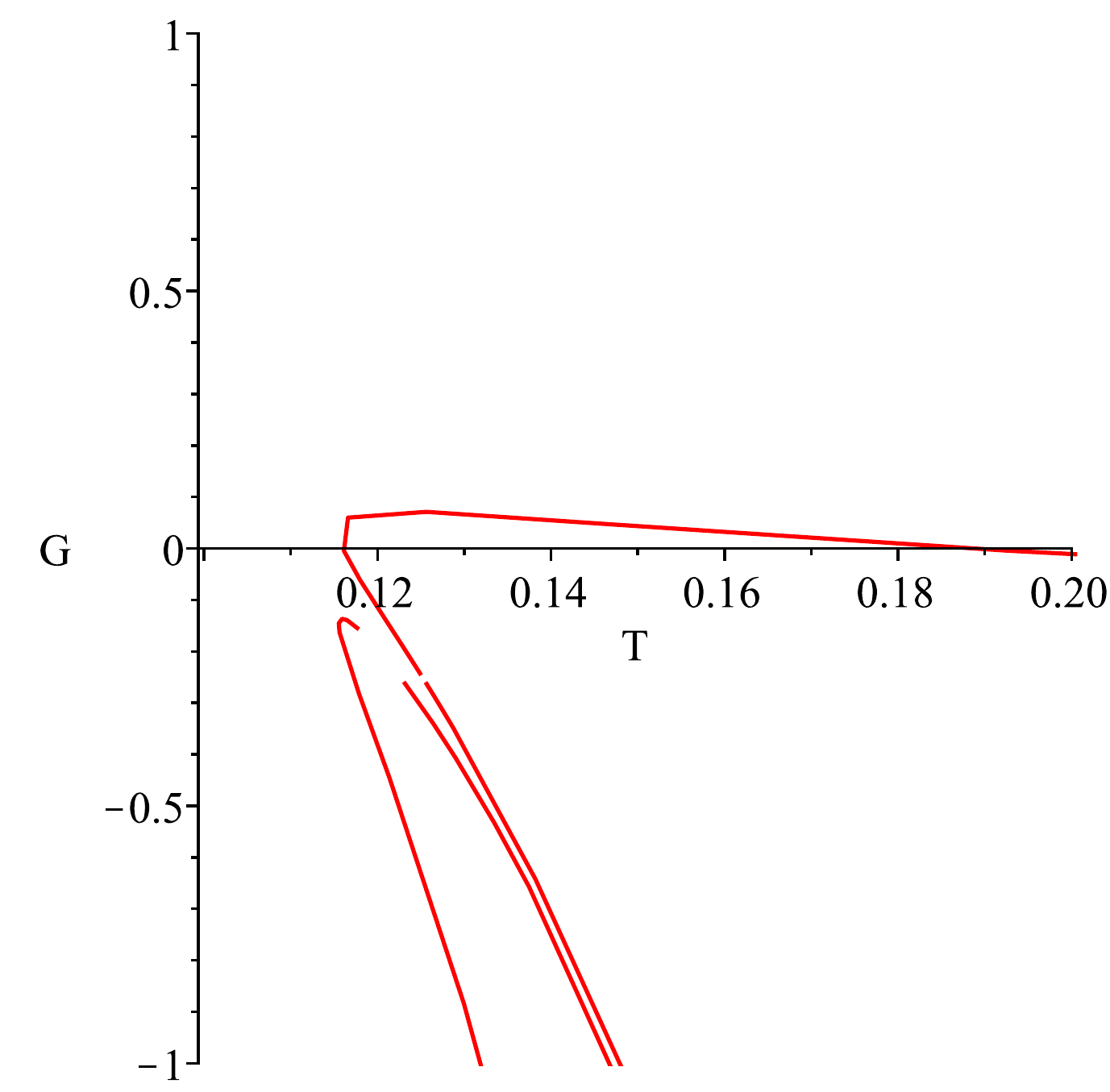} 
    \caption{{\bf Free energy for various  fixed $\psi_{\cal N}$} We set $e=g=0$ and $\ell=3$. 
 The upper blue curve corresponds to    $\psi_{\cal N} = 0$, the lower red curve to $\psi_{\cal N}=0.2$.   The right hand diagram is a close-up of the    $\psi_{\cal N}=0.2$ case, exhibiting the fractured cusp. }
    \label{neutpsi}
\end{figure*} 
 
\begin{figure*}[t!]
    \centering
    \includegraphics[scale = 0.4]{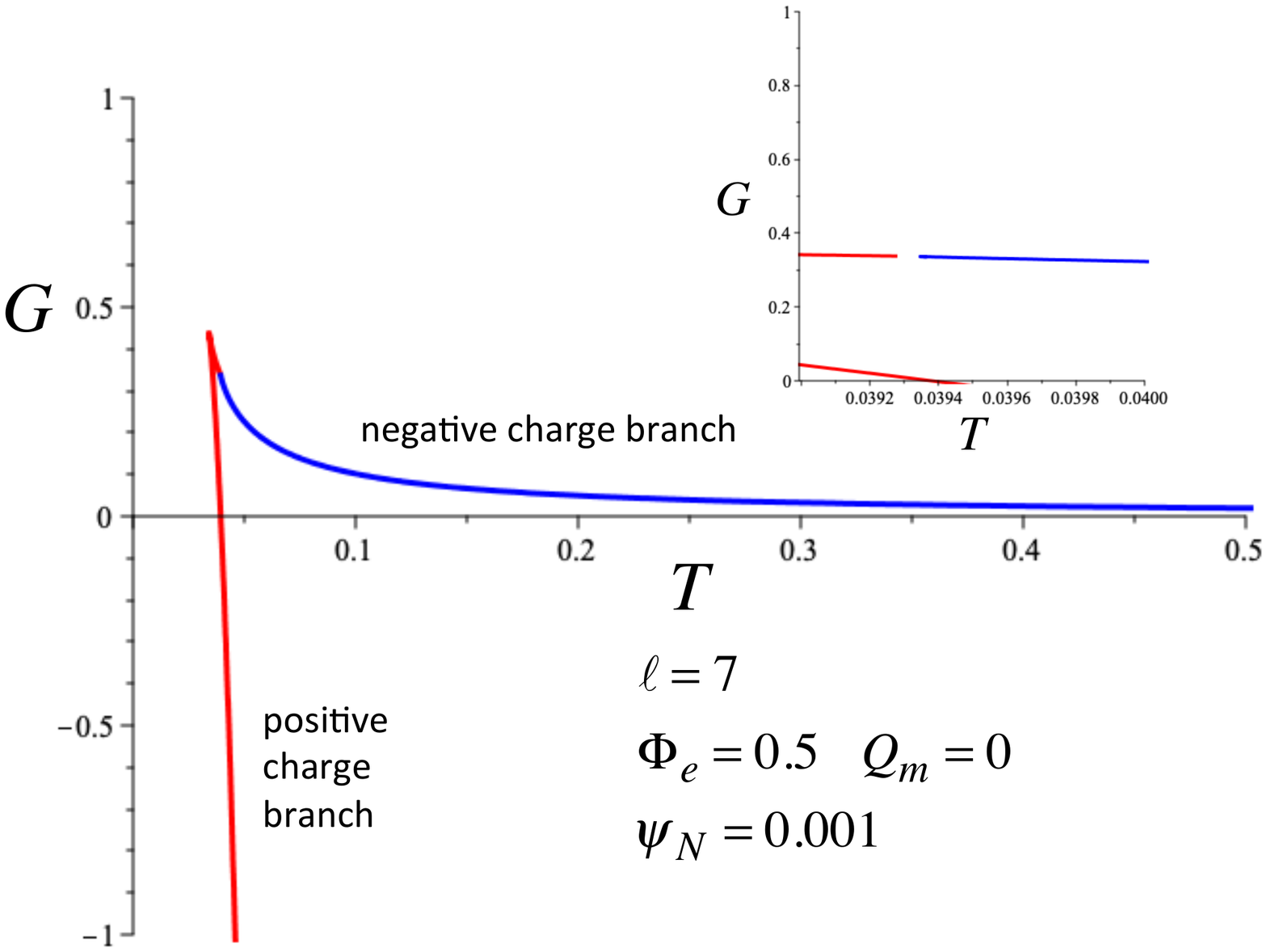} 
    \caption{{\bf Fractured Cusp at fixed electric potential} Free-energy plotted as a function of temperature at vanishing magnetic charge $Q_m=0$, fixed electric potential $\Phi_e = 0.5$, 
    and fixed  $ \psi^{(I)}_{{\cal N}} = 0.001$. The upper blue curve is the negatively charged branch; the fracture is shown in the inset. }
    \label{Fraccusp1}
\end{figure*}

\subsubsection{Fractured Cusp} 

Another  phenomenon is one we refer to as   the `fractured cusp', which can occur for vanishing  magnetic charge, as in cases   1(d) and 2(d). These are equivalent because the electric charge $Q_e$ is the same function of $n$, $r_+$, and $\phi_e$ when $Q_m=0$.  This phenomenon can occur even if $Q_e=0$, as shown in figure~\ref{neutpsi}.

\begin{figure*}[t!]
    \centering
    \includegraphics[scale = 0.45]{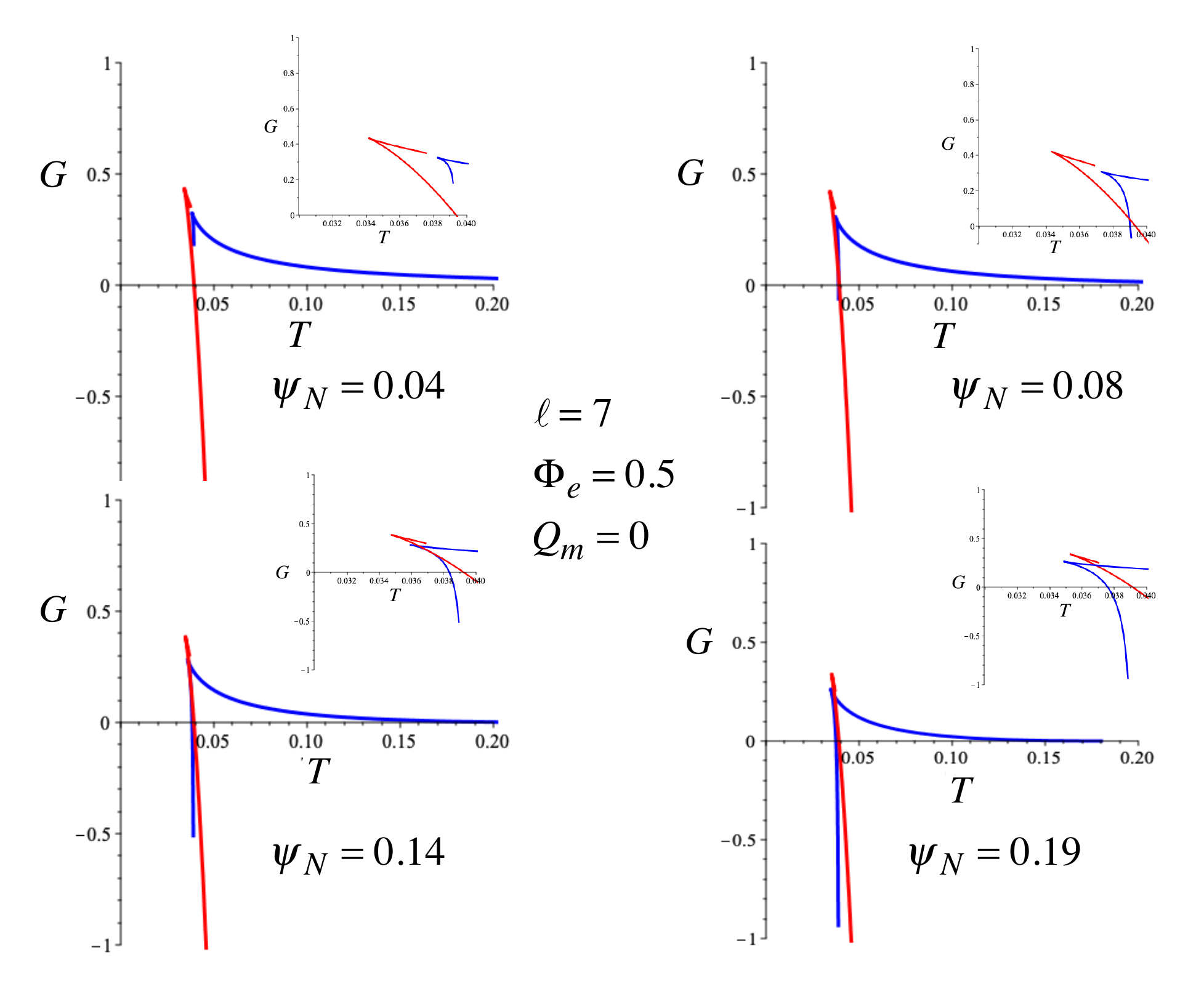} 
    \caption{{\bf Fractured Cusp Development at fixed electric potential}  This sequence of plots of  free-energy vs.  temperature show the development of
    the fractured cusp as $ \psi^{(I)}_{{\cal N}}$ increases for vanishing magnetic charge.   The   blue curve is the negatively charged branch; the insets highlight the relationship    between the negative and positive branches. Here $\ell = 7$ and $\Phi_e = 0.5$.
     }
    \label{Fraccusp2}
\end{figure*}

For vanishing $ \psi^{(I)}_{{\cal N}}$, we observe the familiar cusp structure of a Hawking-Page transition.  Choosing parameters so that $Q_e>0$,
the lower branch of the cusp corresponds to a large black hole of large charge, and the upper branch to a small black hole of small charge.  However if $ \psi^{(I)}_{{\cal N}} \neq 0$,  we find that there is a discontinuity in $Q_e$ as a function of horizon size $r_+$, and that 
black holes of sufficiently small horizon size will be negatively charged, with a corresponding 
 discontinuity appearing  in the temperature.  There is a class of small negatively charged small black holes that is discontinuous from a class of larger positively charged ones, with an intermediate range of $r_+$  where no black holes are possible.   As $ \psi^{(I)}_{{\cal N}} \neq 0$ gets larger, this discontinuity gap widens.  A corresponding discontinuity appears  in the upper branch at small $r_+$ in the free-energy diagram, shown in figure~\ref{Fraccusp1}.
 As $ \psi^{(I)}_{{\cal N}} \neq 0$ increases, this unstable small branch develops a cusp and moves downward to lower values of $G$.  There is a critical value of $ \psi^{(I)}_{{\cal N}} \neq 0$ at which the lower part of the small branch intersects the large branch;  as $ \psi^{(I)}_{{\cal N}} \neq 0$ increases beyond this value we then have a first-order phase transition between a large positively-charged NUT-charged black hole and a small negatively charged one with different NUT parameter.  As $ \psi^{(I)}_{{\cal N}} \neq 0$ becomes even larger, the cusp on the negatively charged branch moves below the positively charged branch, at which point the branch of small negatively charged black holes is thermodynamically stable. This sequence is shown in figure~\ref{Fraccusp2}.
 
\begin{figure*}[t!]
    \centering
    \includegraphics[scale = 0.45]{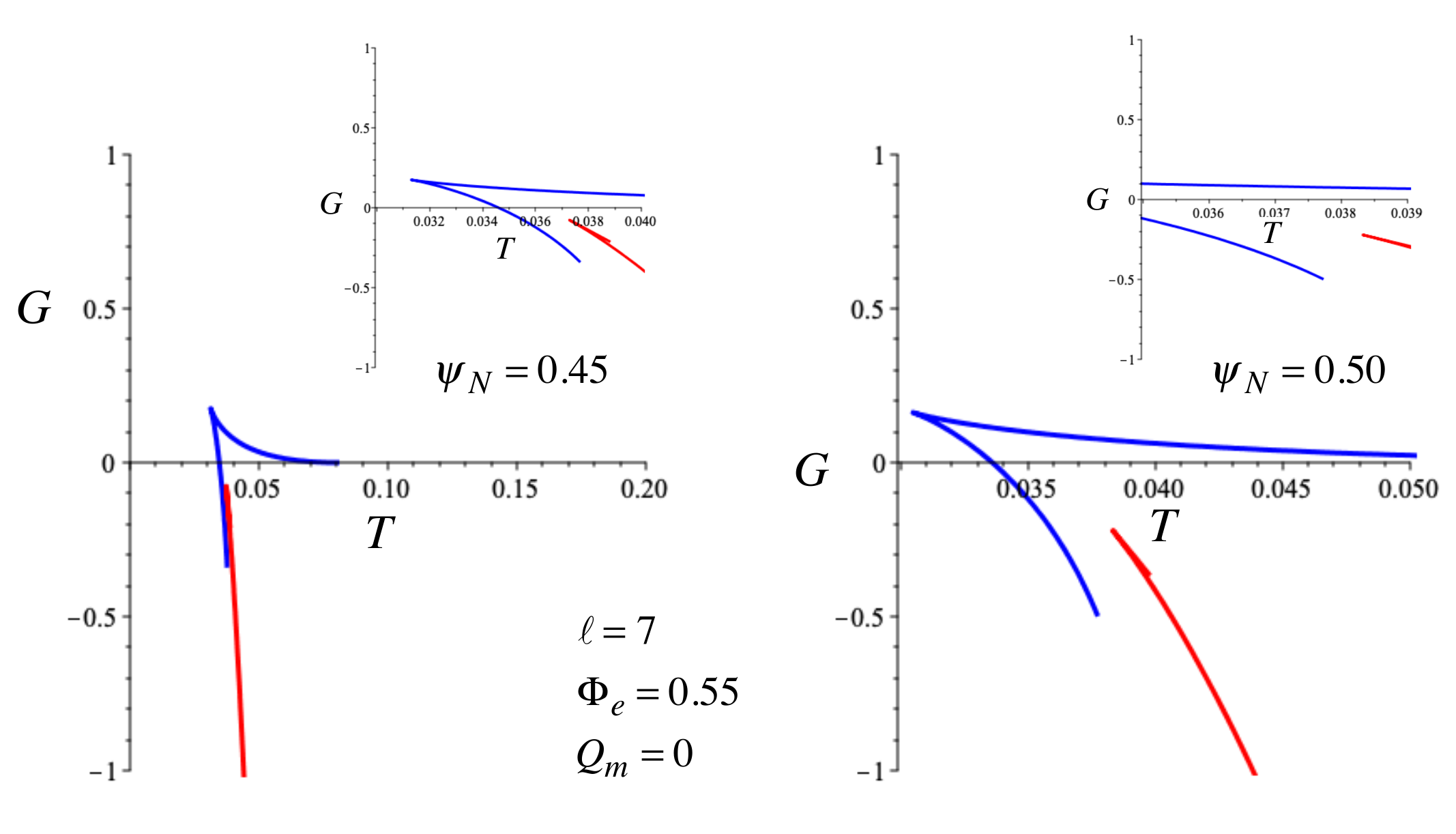} 
    \caption{{\bf Alternate Fractured Cusp Development at fixed electric potential}  This sequence of plots of  free-energy vs.  temperature shows the development of
    the fractured cusp for larger values of $ \psi^{(I)}_{{\cal N}}$ for vanishing magnetic charge.   The   blue curve is the negatively charged branch; the insets highlight the relationship    between the negative and positive branches. Here $\ell = 7$ and $\Phi_e = 0.55$.
     }
    \label{Fraccusp3}
\end{figure*}

 If the electric potential in \eqref{potentials} is fixed to larger values, then the discontinuity for small $ \psi^{(I)}_{{\cal N}} > 0$ is not present, and there is no
 value of the horizon size for which the black hole changes the sign of its charge.  The free energy does not have an unstable branch of small negatively charged black holes; instead the familiar cusp associated with a Hawking-Page transition appears, with the (positively) charged black hole unstable to discharge into thermal AdS.   However once  $ \psi^{(I)}_{{\cal N}} $ is sufficiently large, the unstable branch of small negatively charged black holes reappears, with the value of the horizon radius where the charge becomes negative getting larger for larger   $ \psi^{(I)}_{{\cal N}}$.  As $ \psi^{(I)}_{{\cal N}}$ increases, the  discontinuity
 in the free energy reappears, with a double cusp structure developing.  The right most cusp is the  positively charged branch that is stable for large temperatures, whereas the left cusp is the negatively charged branch stable at smaller temperatures. There is a 0th-order phase transition from a large positively charged black hole to a small negatively charged one, which is stable as temperature decreases until $G=0$, at which point there is the discharge transition to thermal AdS. As 
  $ \psi^{(I)}_{{\cal N}}$ increases further,   a gap between the positive  branch and   negative  branches appears.  For large temperature, the stable branch is of positive charge.  As the temperature decreases,  there is a 0th-order phase transition   to  the upper branch of the negative-charge cusp. As $T$ further decreases, there is another 0th order transition to the lower part of  the negative-charge branch, which is the most stable part.  This situation is illustrated in figure~\ref{Fraccusp3}.

\subsubsection{Snapping Fractured Cusps} 
 
When the magnetic charge $Q_m\neq 0$, new phenomena emerge.  Consider first  $ \psi^{(I)}_{{\cal N}} = 0$. For large $r_+$, the sign of $Q_e$ is reversed relative to the $Q_m=0$ case; for large $r_+$, black holes of large  $Q_e >0 $  become black holes of small $Q_e  < 0 $ for $Q_m > 0$. As $r_+$ gets smaller, there is a critical value at which $Q_e \to -\infty$ for some fixed $Q_m > 0$. Below this value of $r_+$, $Q_e >0$ and very closely matches its value and $Q_m=0$ for a given
value of (small) $r_+$.  There is a corresponding discontinuity in the temperature $T$: as shown in figure~\ref{case1Tvsrp},
below the critical horizon value, $T$ is close to its $Q_m=0$ counterpart, whereas above this value $T$ does not smoothy match this case.    

 \begin{figure*}[h]
    \centering
    \includegraphics[scale = 0.35]{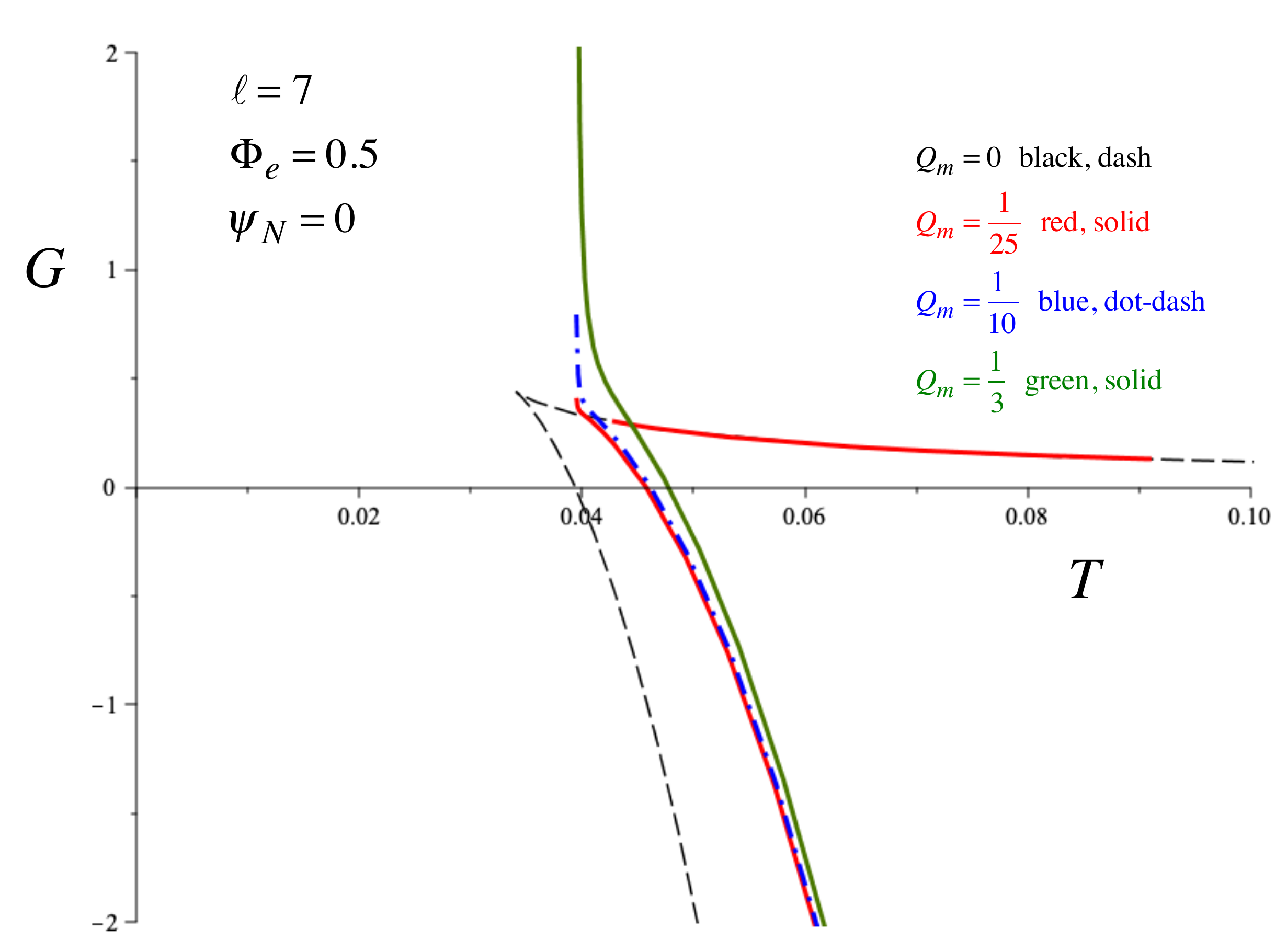} 
    \caption{{\bf Snapping Fractured Cusp}  This structure develops for vanishing $ \psi^{(I)}_{{\cal N}} $ as $Q_m$ increases.  The discontinuity in the fractured cusp widens as    $Q_m$ increases, with the right-hand branch vanishing (snapping away), leaving a single curve that is very steep for small $T$.
     }
    \label{snapcusp}
\end{figure*} 

 \begin{figure*}[h]
    \centering
    \includegraphics[scale = 0.35]{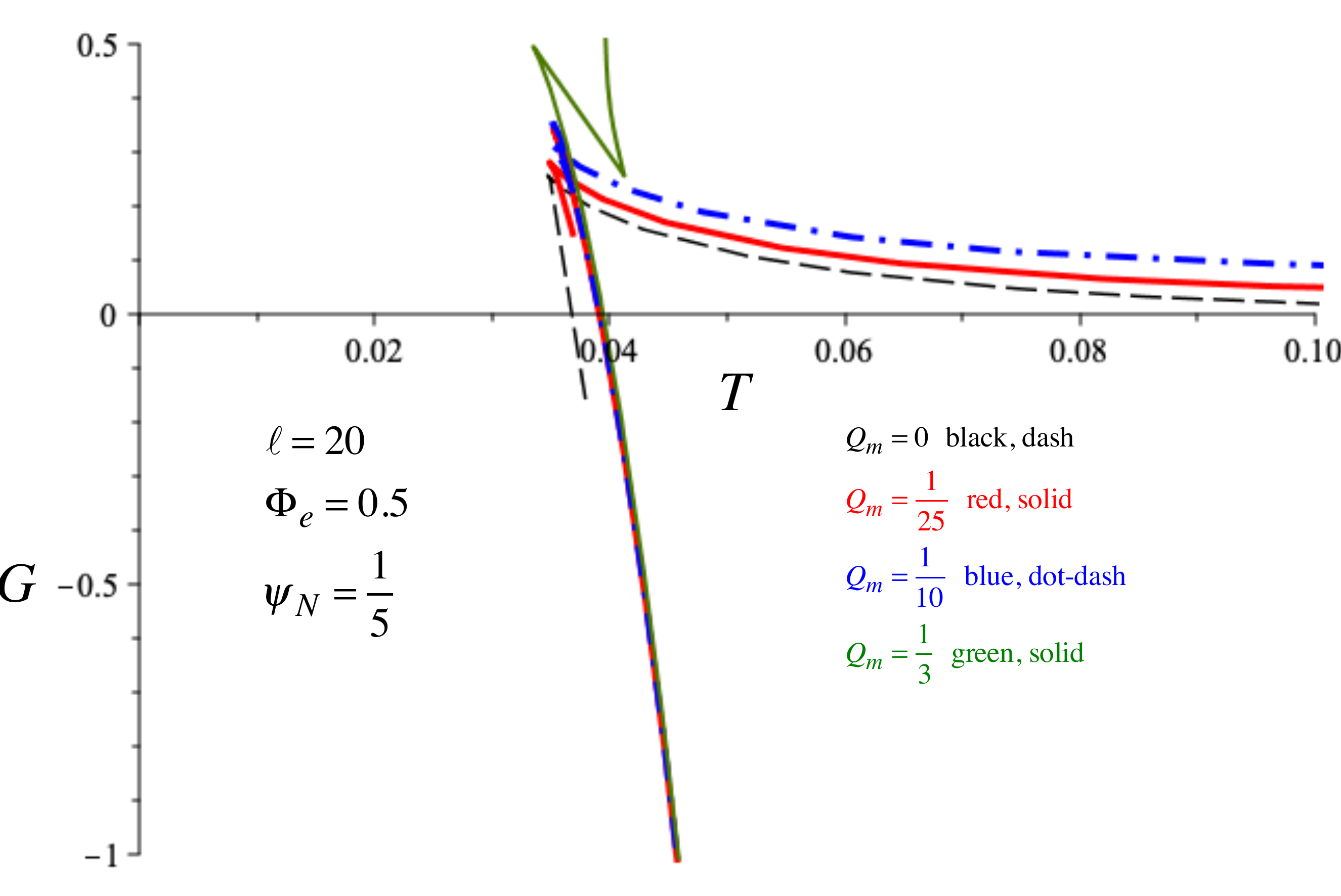} 
    \caption{{\bf Zig-Zag}  This structure develops for nonzero $ \psi^{(I)}_{{\cal N}} $ as $Q_m$ increases.  The discontinuity in the fractured cusp widens as
    $Q_m$ increases, eventually getting connected by a third line and forming a zig-zag structure for sufficiently large $Q_m$.
     }
    \label{zigzag}
\end{figure*} 
 \begin{figure*}[h]
    \centering
    \includegraphics[scale = 0.35]{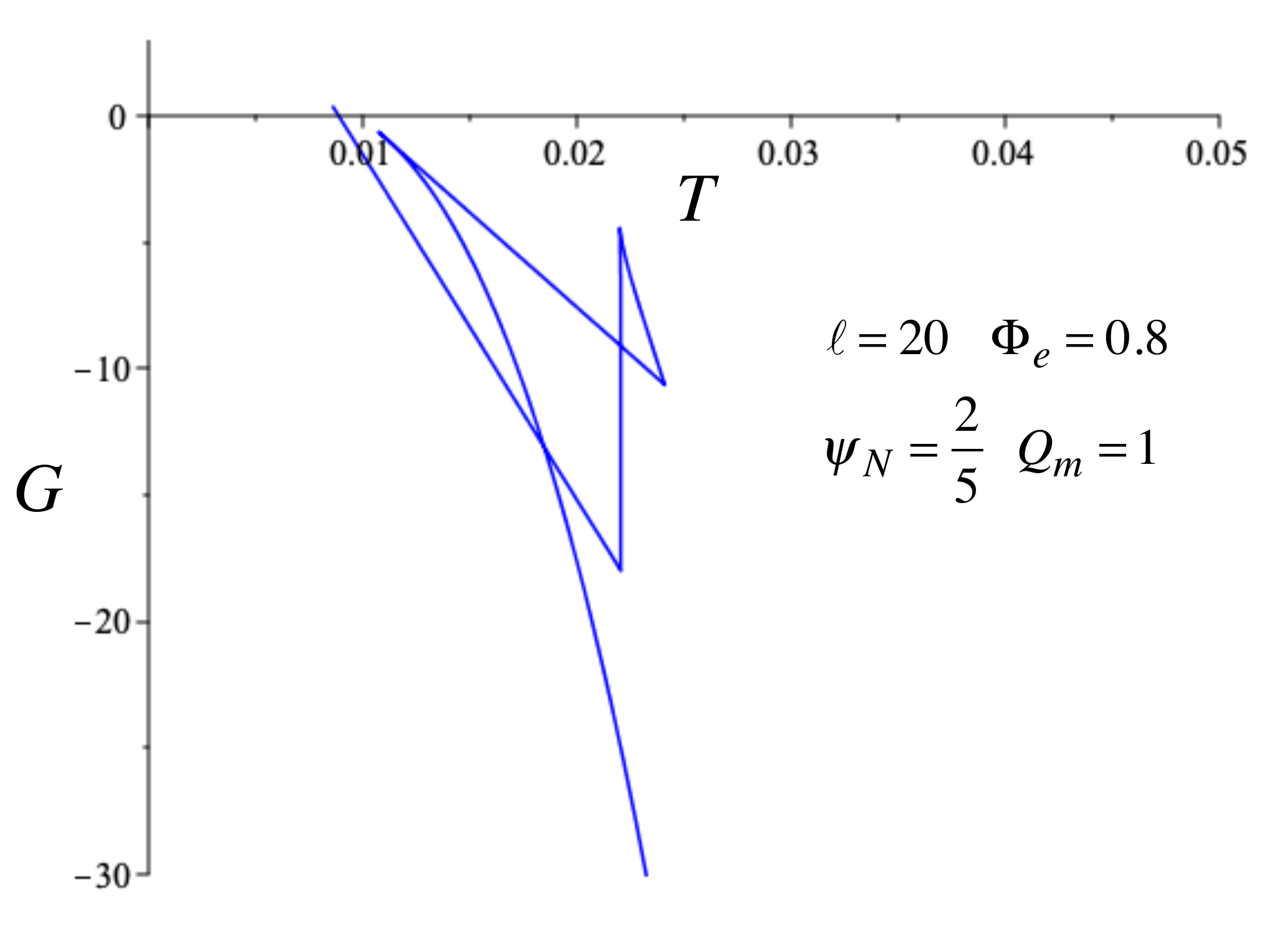} 
    \caption{{\bf Double Swallowtail}  This structure can appear for sufficiently large $\ell$, $ \psi^{(I)}_{{\cal N}} $, and $Q_m$. Large black holes are at the most negative value of the free energy $G$; as temperature decreases there is a first-order large/small phase transition, shown at the most negative self-intersection of $G$.
     }
    \label{dbleswallow}
\end{figure*} 

As noted above, if $Q_m = 0$ and $ \psi^{(I)}_{{\cal N}} > 0$ the sign of $Q_e$ is flipped for small black holes.  But for $Q_m > 0$, the sign of $Q_e$ is retained for both small and sufficiently large black holes, with an intermediate range of $r_+$ where $Q_e$ flips sign. This range gets smaller as  $ \psi^{(I)}_{{\cal N}}$ gets larger. The discontinuity in the temperature is likewise eliminated for large $r_+$, matching closely its $Q_m=0$ counterpart.  There is an intermediate range of $r_+$ where the discontinuity persists; this range increases with increasing $Q_m$ and decreases with increasing $ \psi^{(I)}_{{\cal N}}$.

Turning to the free energy, we find for $ \psi^{(I)}_{{\cal N}} = 0$  that the cusp present in a $G$ vs. $T$ diagram for $Q_m=0$ shifts by a finite rightward amount for small $Q_m\neq 0$. As $Q_m$ gets larger,  the cusp fractures slightly in its upper branch.   As $Q_m$ increases further still, the upper branch of the cusp `snaps' at a threshold value of $Q_m$, becoming a near-vertical steep line. This is the phenomenon of the snapping fractured cusp, shown in figure~\ref{snapcusp}.

For $ \psi^{(I)}_{{\cal N}} > 0$, we find similar behaviour for small $Q_m$ -- the cusp becomes truncated at its lower end, shifting upward and fracturing as $Q_m$ gets larger. However for larger $Q_m$ we encounter a new phenomenon.

 \clearpage

\subsubsection{Zig-zags and Double Swallowtails}

For sufficiently large $ \psi^{(I)}_{{\cal N}} > 0$ and $Q_m$ we observe a new phenomenon.  The discontinuity in the fractured cusp can suddenly become filled, once $Q_m$ exceeds a threshold value.  The $G$ vs. $T$ curve becomes a `zig-zag' structure, shown in in figure~\ref{zigzag}. For a given value of $Q_m$ above this threshold, increasing values of $ \psi^{(I)}_{{\cal N}} > 0$ cause the zig-zag to fold back on itself, producing a double swallowtail structure, shown in figure~\ref{dbleswallow}.  For larger $\ell$ (smaller pressures)  and larger $Q_m$ these double swallowtail structures become more prominent.   They will snap back into a zig-zag structure as $Q_m$ becomes even larger.

We find that all of these phenomena occur at sufficiently small $r_+$. For large enough $r_+$, the $G$ vs. $T$ curves merge, over a broad range of values of $Q_m$ and   $ \psi^{(I)}_{{\cal N}} > 0$.  The phase of the black hole system at a given temperature is, as usual, found from the global minimum of $G$, and phase transitions will occur at all points where $G$ interects itself. For $Q_m\neq 0$, no Hawking-Page transitions will occur as $G$ goes from negative to positive due to the conservation of magnetic charge.
  
 The behavior of $G$ vs $T$ curves for the case II  is qualitatively the same as for  case I. There is an exception for case II, column (d)  and (e) in the table for  Snapping Fractured Cusps.

\section{Concluding remarks}

We have investigated the thermodynamic behaviour of electrically and magnetically charged Lorentzian Taub-NUT black holes, regarding the entropy of these objects to be the Noether charge entropy $S_{\cal N}$
that includes the horizon area and the contribution from the Misner string,  in contrast to the proposal that the only contribution to the entropy comes from the horizon area \cite{Bordo:2019tyh}, denoted $S_+$.  In both approaches the cohomogeneity of the 1st law is the same, and the definitions of the thermodynamic NUT potential $\psi$ and its conjugate charge $N$ are changed.   If $S_{\cal N}$ is taken to be the entropy of the system,
then it can become negative for sufficiently small black holes, and it will diverge (along with $\psi$) 
as the black hole approaches extremality.  However if $S_+$ is taken to be the entropy of the system, then
the $NUT$ potential $\psi$ will diverge as $n\to 0$ (or at some other finite value of $n$ if
we do not take the Misner string  to be symmetrically placed along the polar axis \cite{Bordo:2019tyh}).

 Despite these distinctions, we do not find any other criteria to be significant in distinguishing which choice
of the entropy is preferable.   Indeed, we have seen that charged Lorentzian Taub-NUT black holes exhibit a rich range of unusual phase behaviour, regardless of which interpretation of the entropy is employed.  Columns (a), (c), and (e) of table I yield the various phase behaviours observed in section \ref{FixedN} for both interpretations.  Some of this behaviour was noted previously
\cite{Ballon:2019uha}, but  much is new.   The behaviour seen in section \ref{FixedPsi} applies only if $S_{\cal N}$ is taken to be the entropy of the system, and applies to columns (b), (d) and (e) of table I.  

The physical relevance of Lorentzian NUT charged black holes is still very much an open question in gravitational physics. To the extent that they are relevant, the thermodynamic behaviour we have described will necessarily have to emerge from some deeper quantum gravitational description whose degrees of freedom are not necessarily associated only with the horizon.

\section*{Acknowledgements} This work was supported in part by the Natural Sciences and Engineering Research Council of Canada.

\bibliography{References}


\end{document}